\documentclass[conference]{IEEEtran}
\IEEEoverridecommandlockouts
\usepackage{cite}
\usepackage{algorithmic}
\usepackage{graphicx}
\usepackage{todonotes}
\usepackage{textcomp}
\usepackage{xcolor}
\def\BibTeX{{\rm B\kern-.05em{\sc i\kern-.025em b}\kern-.08em
    T\kern-.1666em\lower.7ex\hbox{E}\kern-.125emX}}

\makeatletter
\newcommand{\@chapapp}{\relax}%
\makeatother

\usepackage{xcolor}
\usepackage{fancyhdr}
\usepackage{subcaption}
\usepackage{listings}
\usepackage{url}

\usepackage[skip=8pt]{caption}

\newcommand{\etal}{\textit{et al}. }

\newcommand{\shorttilde}{{\raise.17ex\hbox{$\scriptstyle\sim$}}}

\newcommand{\img}[4][0.45]{
	\begin{figure}[th]
		\centering { \includegraphics[width=#1 \textwidth]{#2} }
		\vspace{-0.3em}
		\caption{#4} \label{#3}
	\end{figure}
}
\newcommand{\wideimg}[4][0.95]{
	\begin{figure*}[th]
		\centering { \includegraphics[width=#1 \textwidth]{#2} }
		\vspace{-0.3em}
		\caption{#4} \label{#3}
	\end{figure*}
}

\newcommand{\imgstacked}[8]{
\begin{figure}[th]
\centering
\begin{subfigure}[b]{0.45\textwidth}
   \includegraphics[width=1\linewidth]{#1}
   \vspace{-1.5em}
   \subcaption{#2}  \label{#3} 
\end{subfigure}
\begin{subfigure}[b]{0.45\textwidth}
   \includegraphics[width=1\linewidth]{#4}
   \vspace{-1.5em}
   \subcaption{#5}   \label{#6}
\end{subfigure}
\vspace{-0.3em}
\caption{#7} \label{#8}
\end{figure}
}

\usepackage{graphicx}
\begin{document}

\title{Speeding up enclave transitions for IO-intensive applications}

\author{\IEEEauthorblockN{Jakob Svenningsson}
\IEEEauthorblockA{\textit{KTH Royal Institute of Technology} \\
Stockholm, Sweden \\
jaksve@kth.se}
\and
\IEEEauthorblockN{Nicolae Paladi}
\IEEEauthorblockA{\textit{Lund University and CanaryBit AB} \\
Lund, Sweden \\
nicolae.paladi@eit.lth.se}
\and
\IEEEauthorblockN{Arash Vahidi}
\IEEEauthorblockA{\textit{RISE Cybersecurity} \\
Lund, Sweden \\
arash.vahidi@ri.se}
}

\maketitle              

\begin{abstract}
Process-based confidential computing enclaves such as Intel SGX can be used to protect the confidentiality and integrity of workloads, without the overhead of virtualisation. However, they introduce a notable performance overhead, especially when it comes to transitions in and out of the enclave context. Such overhead makes the use of enclaves impractical for running IO-intensive applications, such as network packet processing or biological sequence analysis. We build on earlier approaches to improve the IO performance of work-loads in Intel SGX enclaves and propose the SGX-Bundler library, which helps reduce the cost of both individual single enclave transitions well as of the total number of enclave transitions in trusted applications running in Intel SGX enclaves. We describe the implementation of the SGX-Bundler library, evaluate its performance and demonstrate its practicality using the case study of Open vSwitch, a widely used software switch implementation.
\end{abstract}


\section{Introduction}
\label{sec:introduction}
Confidentiality and integrity are important topics when computation moves from local premises to a third-party environment.
Addressing these topics should not offset the two core advantages of cloud computing: cost reduction and flexibility.
\textit{Confidential computing} is an increasingly popular approach to achieving this~\cite{zhu:2020}.
It relies on using a Trusted Execution Environment (TEE) backed by certified hardware, such that critical operations of Trusted Applications running inside the TEE cannot be manipulated by the platform operator or malicious entities (with the notable exception of the CPU manufacturer).
For example, AMD SEV, Intel SGX, and IBM SVM provide mechanisms to achieve this in different ways~\cite{kaplan2016,Mckeen,yao:2020}.
The variety of vendor TEE implementations highlights trade-offs between security guarantees, portability of legacy applications, ease of deployment, and run-time performance.
VM-based TEE implementations (e.g. AMD SEV, IBM SVM, and Intel TDX) support portability of legacy applications with a modest performance overhead~\cite{gottel:2018}, but have a larger attack surface and are vulnerable to several classes of attacks~\cite{li:2019}.
Process-based TEEs (e.g. Intel SGX and ARM TrustZone) on the other hand have a smaller attack surface and improved security.
Unfortunately, the additional security checks together with memory access limitations also affect the performance of process-based TEEs negatively~\cite{gottel:2018}. 
Furthermore, these have shown to be particularly vulnerable to microarchitectural attacks~\cite{schwarz:2020} and platform vendors have repeatedly issued microcode patches to alleviate security problems~\cite{genkin:2021}. 
Figure~\ref{figure:ecall_transition} illustrates that microcode updates for Intel SGX have caused TEE performance to decrease even further. 

\imgstacked{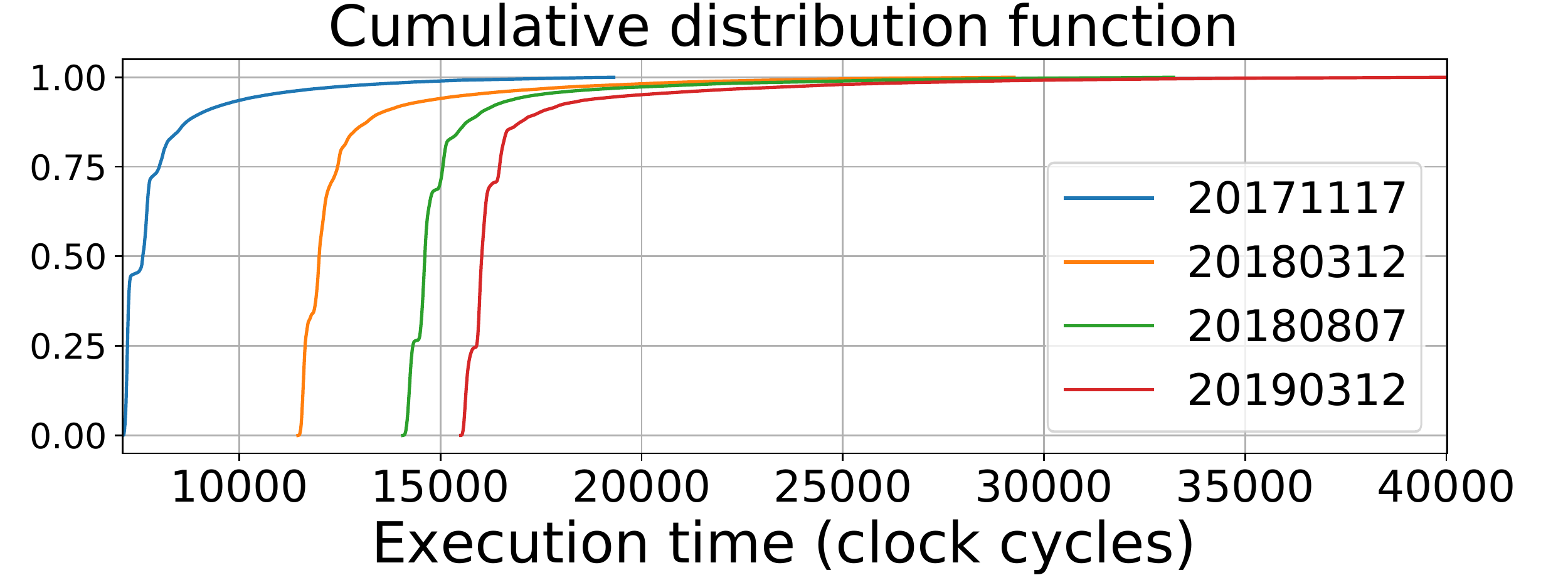}{warm cache}{fig:sub1}{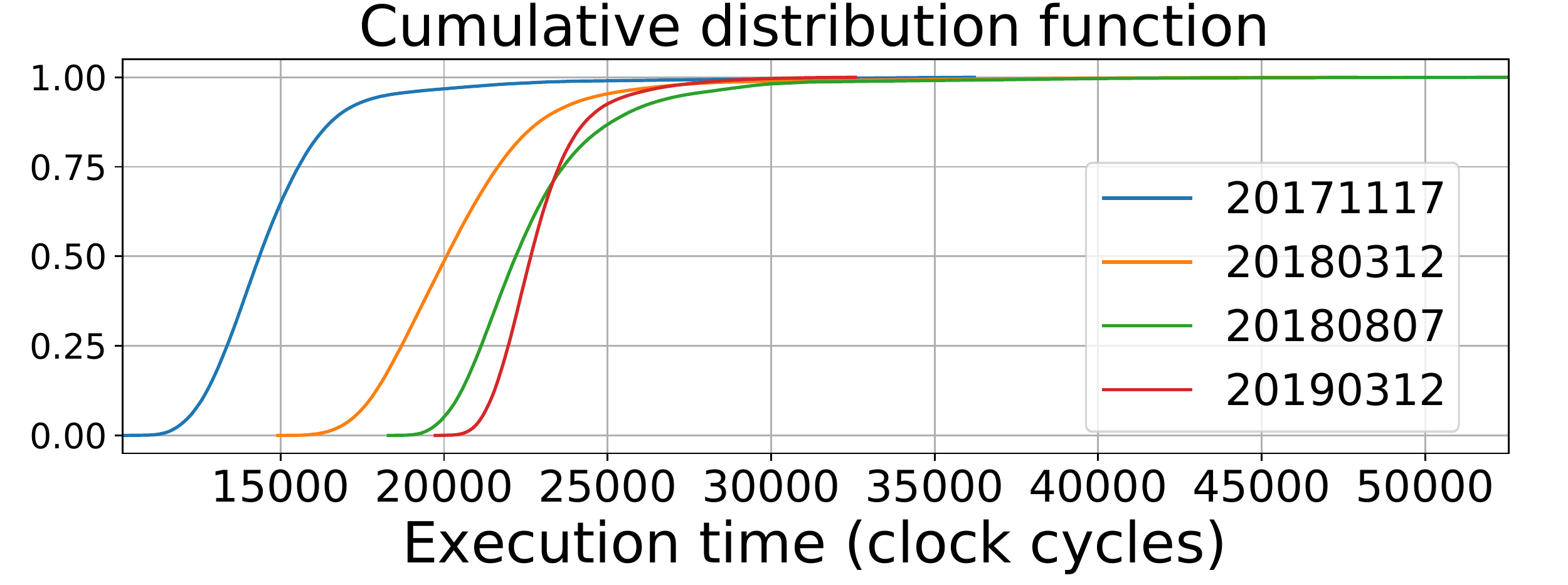}{Cold cache}{fig:sub2}{Evolution of the SGX enclave transition time though Intel microcode updates, presented as a cumulative distribution function (CDF).}{figure:ecall_transition}

It is therefore imperative to identify and implement new approaches that help to maintain or improve TEE performance despite the latest countermeasures to microarchitectural attacks. 
This, however, must also be done in a way that does not significantly increase application developers' efforts. 
In this paper, we address a crucial limitation on the intersection between the portability of legacy applications and the performance overhead introduced by the transition between the TEE and the Rich Execution Environment (REE).

Our results show that while a tailor-made refactoring of legacy Trusted Applications for  Intel SGX yields the best performance, it is labor-intensive, application-specific, and often impractical.
This insight led us to develop the \textit{SGX-Bundler} software library as a generic approach for speeding up enclave transitions in legacy Trusted Applications, while maintaining the security benefits of SGX.
This solution is particularly beneficial in IO-intensive applications such as network packet processing, remote sensing applications, biological sequence analysis, and long-running simulations~\cite{qin:2009}. 
We demonstrate the practical applicability and performance improvements of our approach using a packet processing application. Note however that the proposed solution is generic and can be applied to any domain.
This paper extends, clarifies and complements the preliminary results presented in~\cite{svenningsson:2021}.
The main \textbf{contributions} of our work are summarized as follows:
\begin{itemize}
    \item We describe a generic approach to speed up transitions between the rich execution environment and SGX enclaves (Section~\ref{section:hotcall-bundler-library});
    \item We introduce \textit{enclave execution graphs}, that allow executing arbitrary sequences of enclave functions using a single enclave transition (Section~\ref{section:hotcall-bundler-implementation});
    \item We implement a library to assist refactoring of legacy applications and introduce efficient transitions in and out of SGX enclaves;
    \item We demonstrate the applicability of our approach with the case study of a widely used IO-intensive application;
    \item The implementation source code is openly available\footnote{Source code repository:~\url{https://github.com/nicopal/sgx_bundler}
    }.
\end{itemize}

The rest of the paper is structured as follows.
We introduce the required background and motivate the problem in Section~\ref{section:background}, introduce the SGX-Bundler library in Section~\ref{section:hotcall-bundler-library} and describe the implementation of the library in Section~\ref{section:hotcall-bundler-implementation}.
Next, we evaluate the performance of the SGX-Bundler library and its application in a case study in Section~\ref{section:results}, present the related work in Section~\ref{section:related-work} followed by conclusion and future work in Section~\ref{section:conclusions}.

\section{Background}
\label{section:background}
We next introduce several key concepts used in the paper. 

\subsection{Intel SGX}
\label{subsec:sgx}
Intel Software Guard Extensions (SGX) are CPU security extensions that allow execution of unprivileged trusted applications in the presence of possibly malicious privileged software such as a compromised OS or hypervisor~\cite{Mckeen}. 
An SGX-enabled CPU maintains an isolated memory region, the Enclave Page Cache (EPC), within which security enclaves can execute isolated from the rest of the system. 
SGX provides mechanisms to verify the integrity of an enclave (using local and remote attestation) and binding of information to specific configurations (sealing), which allows one to validate an enclave without direct access to its content.

Enclaves communicate with applications running in the Rich Execution Environment (REE) using the ECALL and OCALL (entry and out call) instructions.
However, these instructions introduce a performance overhead that often makes SGX unsuitable for IO-intensive applications.
Weisse proposed using a shared memory region outside the enclave for communication, resulting in significant performance improvements in real-world applications~\cite{Weisse2017}.
In response to published security vulnerabilities affecting Intel SGX~\cite{wang:2017}, ~\cite{lipp:2018}, ~\cite{kocher:2019}, Intel issued a number of microcode updates.
However, along with addressing software vulnerabilities this further degraded the performance of enclave transitions (see Figure~\ref{figure:ecall_transition}).
While the HotCalls approach~\cite{Weisse2017} produces a tangible performance improvement, we note the importance of further efforts to offset the overhead introduced by subsequent microcode updates.

\subsection{Memoisation}
\label{subsec:memoisation}
Memoization is an optimization technique for reducing the execution time of computationally expensive functions~\cite{Kleinberg2005}.
Given a function with no side effects, memoization uses a cache to remember some input-output pairs.
If an input used in a subsequent call is found in the cache, the recorded output value is returned, otherwise, the (expensive) function call is taken.
Memoization is a simple way of trading execution time for space and is commonly used to optimize recursive algorithms.
We use memoization to speed up enclave transition times between applications running in the TEE and the REE.

\subsection{Open vSwitch}
\label{subsec:openvswitch}

The motivating use case for this work is Open VSwitch (OvS), a software network switch for connecting physical and virtual network interfaces in a virtualized environment~\cite{koponen2014}. This is a critical component for providing network isolation in cloud infrastructure and other multi-tenant environment~\cite{pfaff2009}. 

\img[0.5]{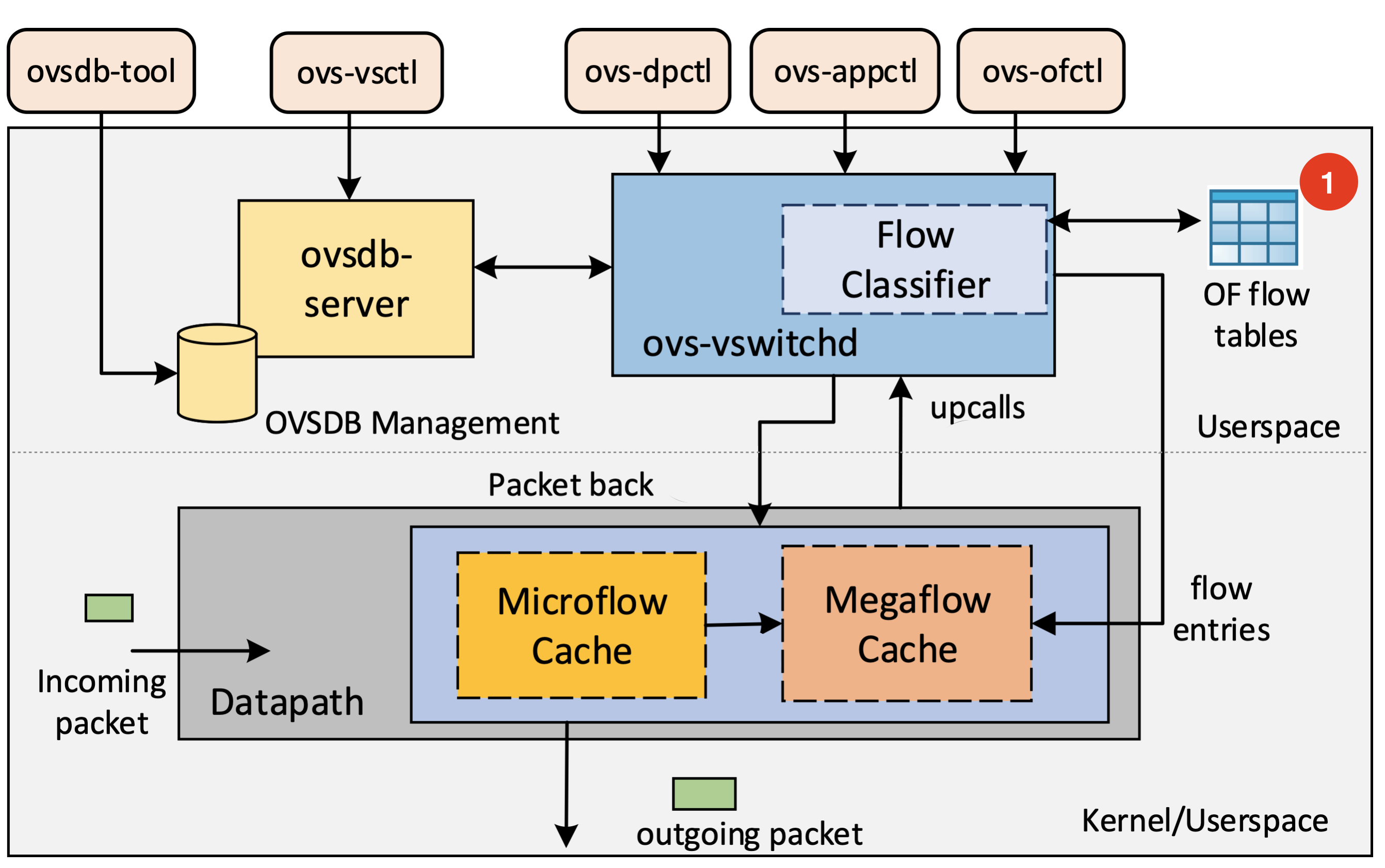}{figure:ovscomponents}{Overview of OvS main components.} 
Among the OvS components (Figure~\ref{figure:ovscomponents}), the \emph{flow tables} (1) are of special interest to us as they contain the rules that define the switch routing behavior. While these tables are critical OvS assets, they are often stored without sufficient confidentiality and integrity protection, leading to serious security vulnerabilities.
For example, an attacker with access to flow tables could map the network structure~\cite{bifulco:2015}, modify routing behavior to perform man-in-the-middle attacks, or bypass firewalls and intrusion-detection systems~\cite{antikainen:2014}. Furthermore, an attacker could inject malicious data into flow tables to propagate deeper into the network and compromise systems otherwise not reachable~\cite{Hong_2015}.

Proposed solutions to address flow table security issues include auditing flow table to detect discrepancies between the configured and current behavior~\cite{10.1145/2857705.2857721}, validating both executables and flow tables with a TPM~\cite{Jacquin2015}, or moving critical components (the OpenFlow flow tables and forwarding logic) into Intel SGX enclaves~\cite{medina:2019}. The latter, while promising from a security point of view, is a very labor-intensive task and introduces additional overhead. In this work, we address both shortcomings.

\subsection{Threat model}
\label{subsec:threat-model}
We focus on the integrity of critical components in applications executing in multi-tenant environments. We assume the critical components execute in TEEs and communicate at high frequency with the corresponding applications in REE.
We consider an adversary capable of operating arbitrary software components and having remote execution capabilities on platforms where target applications operate. The adversary may modify any REE software component.
We exclude microarchitectural attacks~\cite{genkin:2021} and address them in upcoming work; we consider existing countermeasures against such attacks in our performance analysis.

\section{Speeding up Enclave Transitions}
\label{section:hotcall-bundler-library}
We describe \textit{SGX-Bundler}, a mechanism addressing performance penalties caused by transitions from and to SGX enclaves.
To help adoption and usability, we implemented the \textit{SGX-Bundler} library. 

\subsection{Overview}
\label{section:bundler}
The \textit{SGX-Bundler} library offers functionality to reduce the cost of individual enclave transition as well as the total number of enclave transitions for trusted applications (TAs) deployed in Intel SGX enclaves.
This library extends work conducted in HotCalls~\cite{Weisse2017} with novel ideas and is the core contribution of this paper.
The library leverages three main features: switchless enclave function calls, execution graphs, and enclave function memoization. 

Switchless enclave function calls are used to reduce the cost of a single enclave transition. 
Execution graphs and enclave function memoization are used to reduce the total number of enclave function calls in Intel SGX applications.

\subsection{Functional Requirements}
\label{section:implementation-func_req}

We consider the following functional requirements for the SGX-Bundler library, defined based on observations of the performance analysis described in Section~\ref{section:results-ovs}:
(1)~\textbf{Switchless calls}: execute enclave functions without context-switching to enclave mode; 
(2)~\textbf{Merging}: execute an arbitrary number of enclave functions over a single enclave transition;
(3)~\textbf{Batching}: apply an arbitrary number of enclave functions to each element of an input list over a single enclave transition;
(4)~\textbf{Branching}: conditional execution of enclave functions over a single enclave transition;
(5)~\textbf{Memoization}: cache enclave data in untrusted memory when confidentially is not required. 
Caches allow untrusted applications data access without enclave transitions. 
Moreover, we implement a mechanism to verify the integrity of enclave data stored in untrusted memory.

The switchless enclave function call component presented in \ref{section:implementation-hotcall} fullfills requirement 1; the execution graph component in Section \ref{section:implementation-execution_graph} fullfills requirements 2-4, and the memoization component in Section \ref{section:implementation-memo} fullfills requirement 5.

\subsection{Architecture}
\label{section:implementation-arch}

In the case of SGX enclaves, implementing a shared memory switchless enclave communication library requires source code modifications in both the trusted application running in the TEE and the untrusted application running in the REE. 
Enclaves do not share source code (and libraries) with the untrusted application; 
therefore, the SGX-Bundler library consists of two separate libraries. 
The first library is a \textit{static C} library that needs to be linked with the untrusted application, and the second is a trusted enclave library which needs to be linked with the enclave. 
Trusted enclave libraries are static libraries that are linked with the enclave binary \cite{sgx_reference_manual}.

\img[0.5]{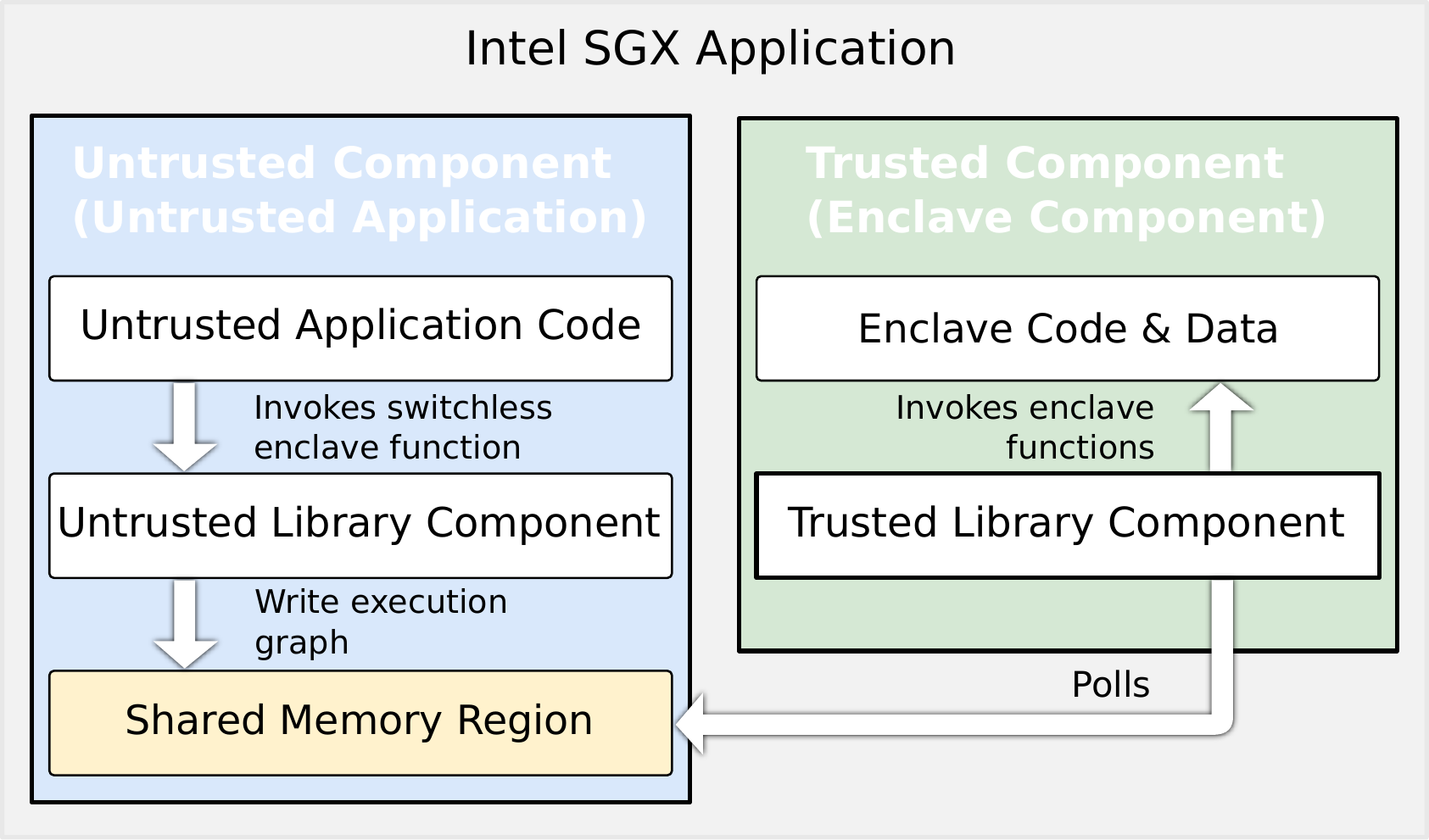}{figure:overview_arch}{High-level overview of an Intel SGX application using the SGX-Bundler library.} 

Figure~\ref{figure:overview_arch} illustrates the untrusted and trusted part of the SGX-Bundler library when integrated into an arbitrary Intel SGX application and the interactions between the different parts.
The untrusted application invokes switchless enclave functions through an API exposed by the untrusted library. 
Next, the untrusted library writes the job to a shared memory region in the form of an execution graph (execution graphs are discussed later in Section \ref{section:implementation-execution_graph}). 
Finally, the job is processed by an enclave worker thread which calls the associated enclave function and writes back potential return values to the shared memory region.

\section{SGX-Bundler Implementation}
\label{section:hotcall-bundler-implementation}
We next describe the SGX-Bundler implementation.

\subsection{Switchless Enclave Function Calls}
\label{section:implementation-hotcall}
The protocol used for switchless enclave function calls in the SGX-Bundler library builds on HotCalls \cite{Weisse2017} and is presented in Figure~\ref{figure:arch}. 
This component fulfills functional requirement (1) listed above in~\ref{section:implementation-func_req}.
The shared memory region contains a \textit{spinlock} primitive that must be acquired by either the untrusted application and the TA before accessing the shared memory region to avoid data races.
While Intel SGX SDK supports condition variables, this synchronization primitive is implemented with OCALLS, which is a context switch operation and conflicts with our goal to keep the communication protocol switchless. 
Spinlock is the only synchronization primitive that can be used by the enclave worker threads without leaving the enclave.

\wideimg{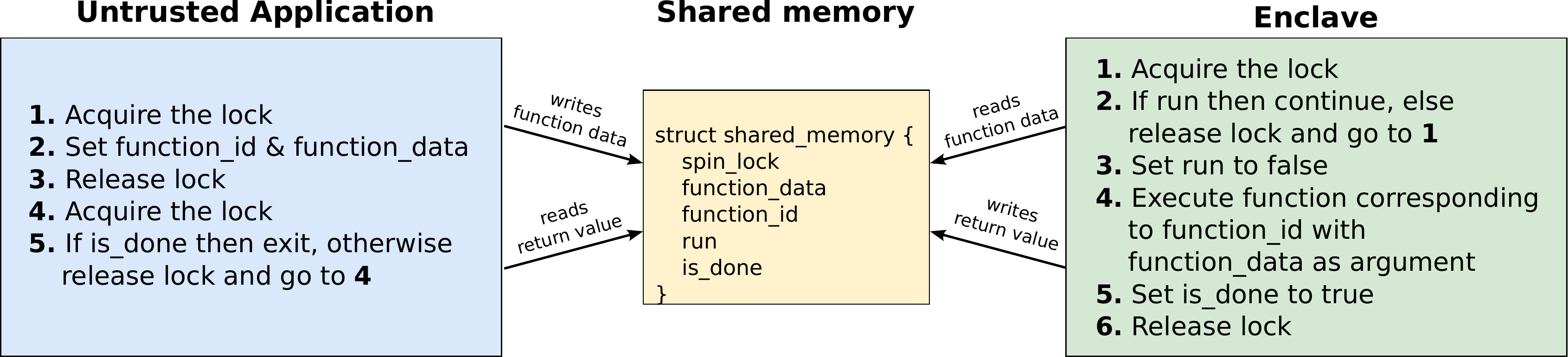}{figure:arch}{Switchless enclave function call protocol.}

The untrusted application invokes switchless enclave functions by acquiring the lock of the shared memory region and writing the enclave function call, represented by a (\textit{function\_id}, \textit{function\_data}) tuple to shared memory. 
An enclave worker thread initiated through an API exposed by the trusted part of the library is continuously polling the shared memory region for scheduled jobs to execute. 
The enclave worker thread uses a busy-waiting scheme where it repeatedly checks for pending jobs inside of an infinite loop.
We use Intel's \textit{pause} instruction inside of the spinlock loop to improve the efficiency of the busy-waiting scheme.
The pause instruction provides a hint to the processor that it is executing inside a spinlock loop, enabling the processor to perform memory optimizations~\cite{intel_pause}. 

In Section \ref{section:implementation-execution_graph} we will replace this tuple with a data structure representing an \textit{execution graph} to create a more efficient enclave communication scheme able to execute multiple enclave functions using a single enclave transition.

\subsubsection{Translation Functions}\label{section:wrapper}
Input and output parameters are treated as generic elements which simplifies the implementation but must be translated to correct data types before an enclave worker thread can be invoked. This is done by defining a translation function for each function exposed to the untrusted application, see Listing \ref{listing:wrapper} for an example.
Note that translation functions are constructed to accept an array of parameters, which will enable the use of batching (see Section \ref{section:implementation-iterator}).

\begin{lstlisting}[
basicstyle=\footnotesize, 
language=C, 
label={listing:wrapper},
caption=A translation function for an enclave summation.]

void translation_ecall_plus( unsigned int itrs, 
  unsigned int params, void *args[][]) {
  for(int i = 0; i < iters; ++i) {
    *(int *) args[2][i] = hotcall_plus(
        *(int *) args[0][i], *(int *) args[1][i]);
  } 
}
\end{lstlisting}

\subsection{Execution Graphs}
\label{section:implementation-execution_graph}

A limitation of the \textit{HotCall} implementation~\cite{Weisse2017} is that it only allows execution of a single enclave function per enclave transition. A switchless enclave transition still introduces an overhead estimated to be around \shorttilde600 to \shorttilde1400 clock cycles for warm and cold caches respectively~\cite{Weisse2017}. A simple approach to address this is to merge sequence of enclave calls into a single call, as illustrated in Figure~\ref{figure:merge}.


\imgstacked{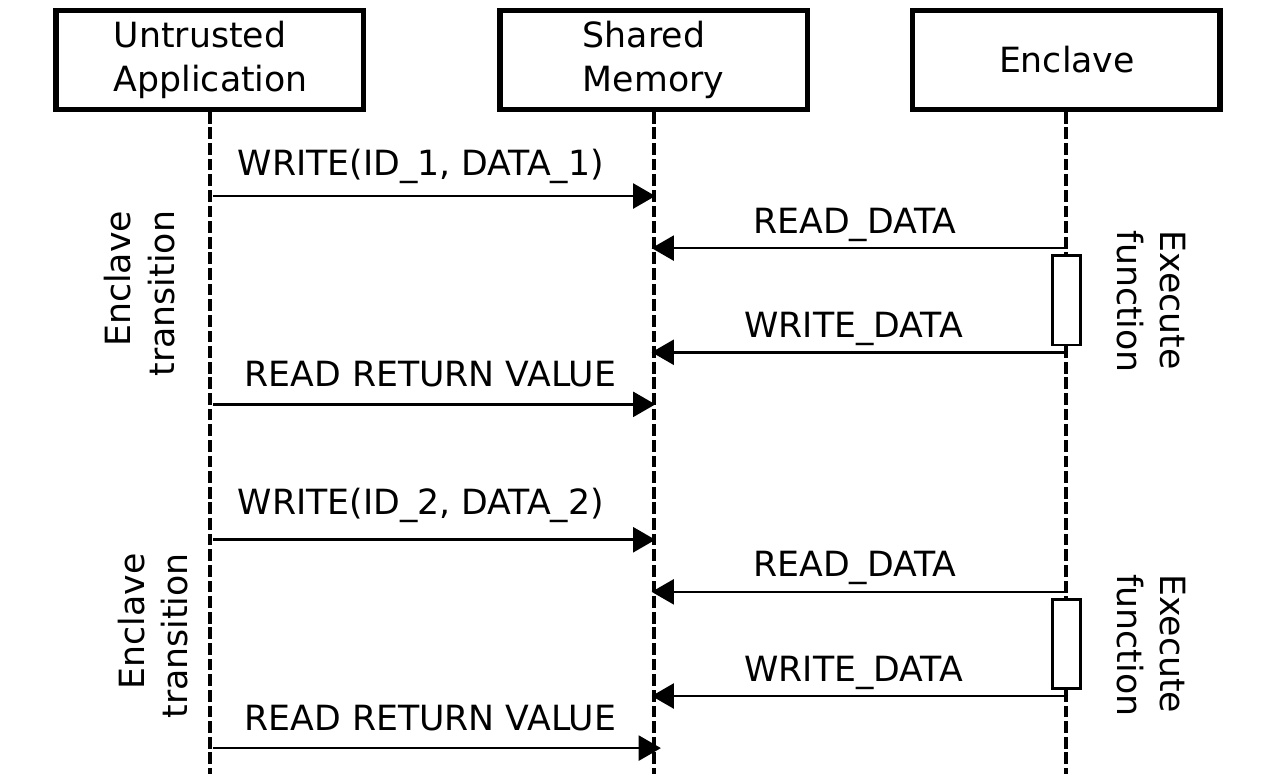}{Two enclave calls}{figure:merge_graph}{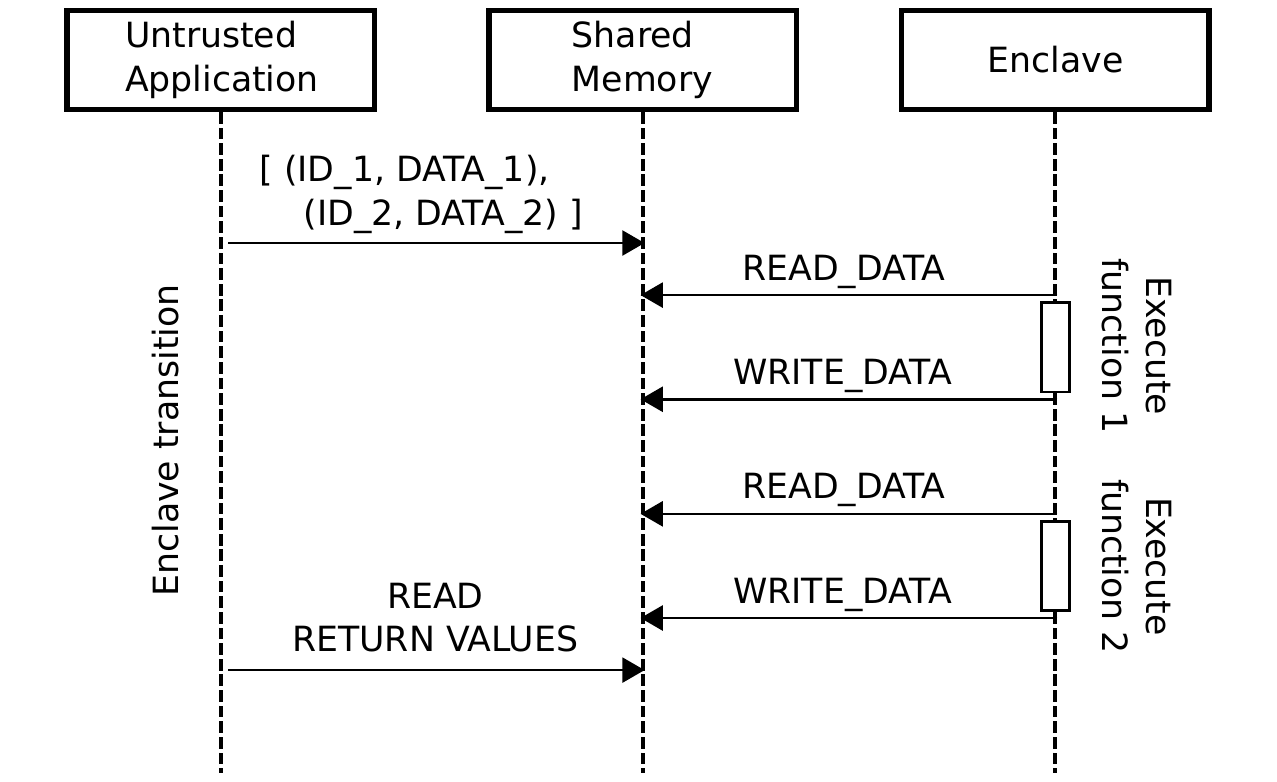}{Two merged enclave calls}{figure:merge_exec_graph}{
Sequence diagram illustrating two function calls with the original HotCall implementation and using execution graphs.}{figure:merge}


In practice the enclave call sequence may be much more complex than just a pre-defined list of function calls. To address this, we introduce the concept of \textit{execution graphs} in the context of enclave transitions. An enclave \textit{execution graph} is an arbitrary sequence of dependent or independent enclave function calls, control statements, and iterators that are executed within a single enclave transition. This provides a significant improvement over the original HotCall implementation and is to best of our knowledge a novel concept that has not been explored in previous studies.
We discuss various graph components and their function outlines in Appendix~\ref{appendix:executiongraphs}.

\subsection{Construction of Execution Graphs}
When converting an imperative programming language to execution graphs, each node can require $5-10$ lines of boilerplate code.
This is a tedious and error-prone task and most likely will result in a less readable code.
To address this problem, we created a user-friendly API based on C pre-processor macros.
This API can be used for building execution graphs using both an imperative and functional-style syntax, and is explained in detail in 
Appendix \ref{appendix:bundler-api}.

\subsection{Enclave Function Memoization}
\label{section:implementation-memo}

While execution graphs are effective capturing complex operations that have a high number of enclave calls, they are not as effective in handling  simpler enclave operations. To address this we make propose caching results of frequent enclave calls in untrusted shared memory using a technique called \textit{memoization}. The integrity of memoization caches in untrusted memory is guaranteed by storing a hash of each memoization cache in the enclave. We compute the hash of a memoization cache as follows:

\begin{equation}
\sum_{e \in C}^{} hash(e)
\end{equation}
where $C$ is the set of all entries in the cache. 
The enclave worker thread, responsible for populating memoization caches, updates the corresponding memoization hash each time a cache entry is inserted or deleted. 

The enclave worker thread periodically verifies the caches by recalculating the hashes of the memoization caches in untrusted memory and compares them with the hashes stored in enclave memory. Depending on the nature of the application, different actions can be appropriate when an unauthorized modification is detected.
Manipulating the eviction list only enables an attacker to give cache priority to selected entries.

\section{Results}
\label{section:results}

To assess the performance gains of using SGX-Bundler, we first perform several micro-benchmarks for each component of the library. To evaluate real-world performance improvements in a significantly more complex environment, we also evaluate a prototype implementation of OvS with SGX support.

\subsection{SGX-Bundler Library}
\label{section:results:bundler}

The following components of the proposed solution are evaluated here: switchless enclave functions, execution graphs (merging, batching, branching), and memoization. This allows us to study the benefits of each improvement in isolation.

\subsubsection{Enclave Transition Time}
\label{section:switchless}
\imgstacked{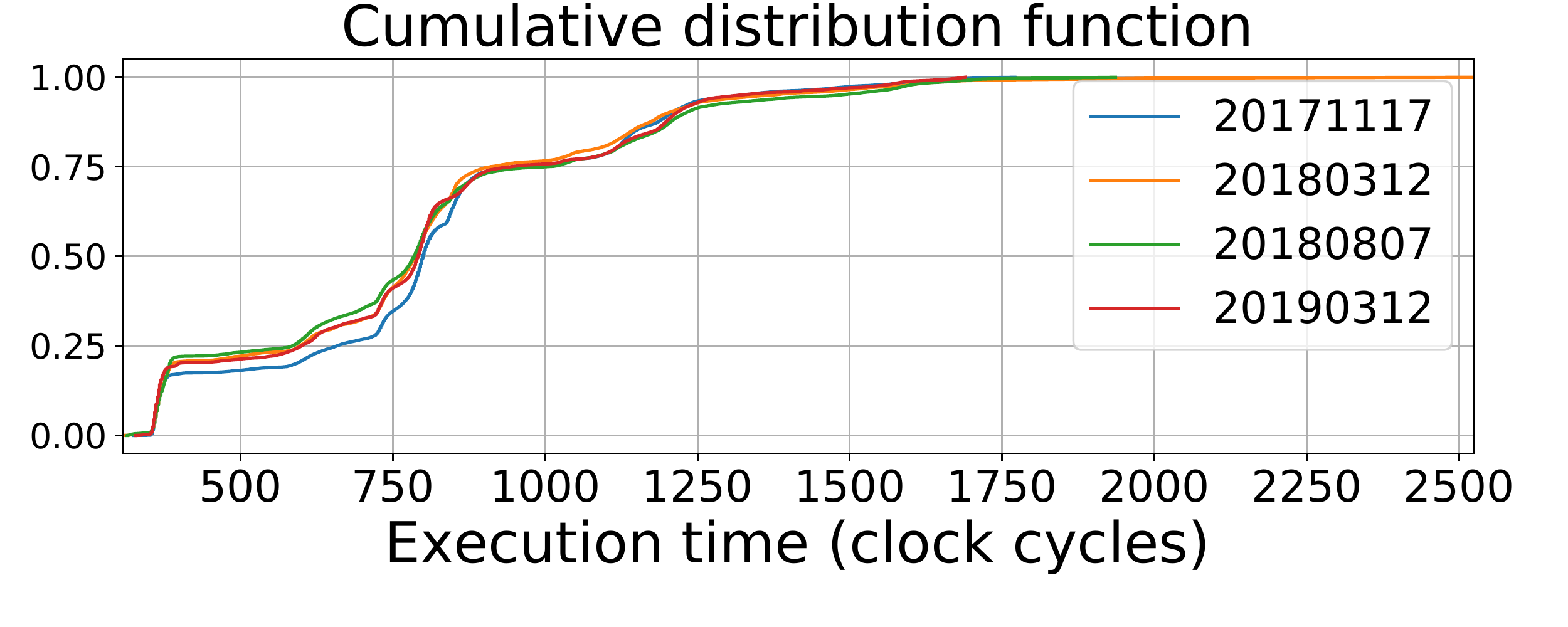}{warm cache}{fig:subfigure1}{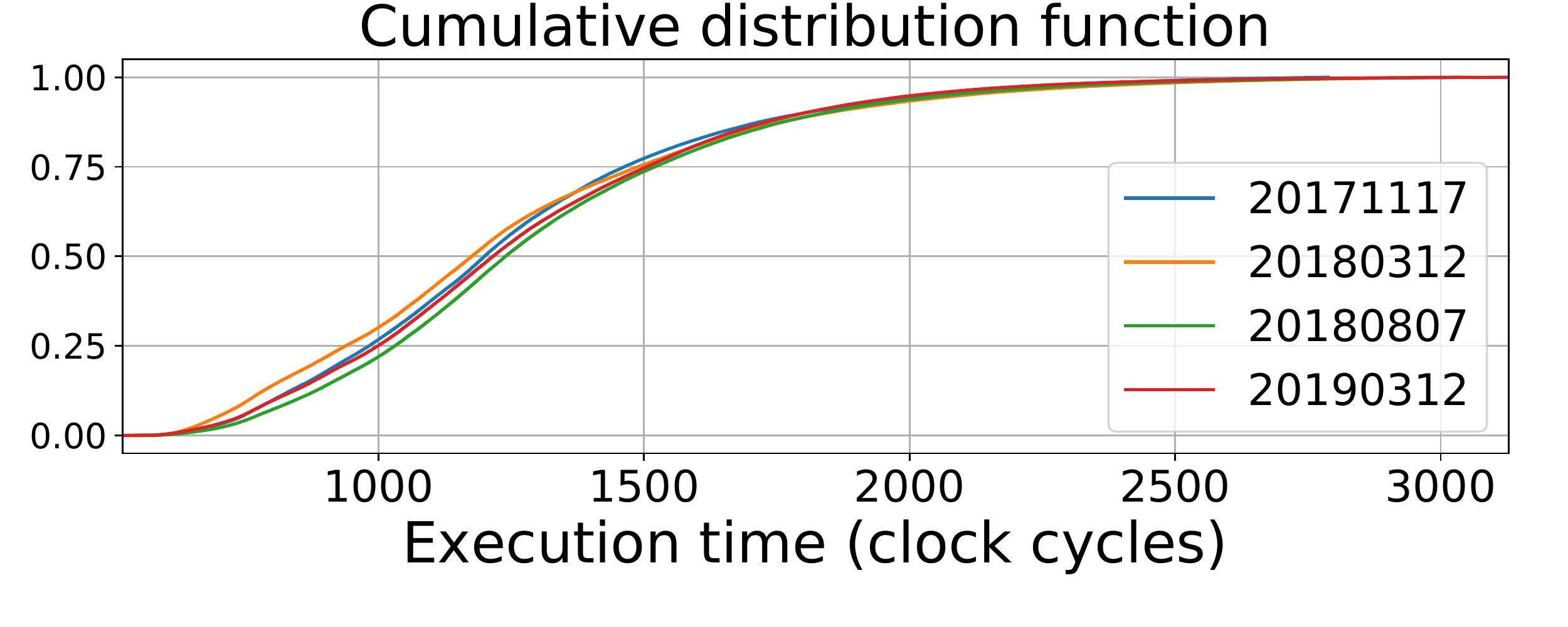}{cold cache}{fig:subfigure2}{Enclave transition times for switchless enclave function calls with different Intel microcode versions.}{figure:hotcall_transition}

Measured execution time for switchless function calls can be observed in Figure~\ref{figure:hotcall_transition}. Compared to ECALLs (see Figure~\ref{figure:ecall_transition}), we note that not only the switchless calls are significantly faster (\shorttilde\textbf{20.3x} and \shorttilde\textbf{18.6x} for warm and cold cache), they are also mostly unaffected by microcode updates. In contrast, ECALL warm and cold cache performance has increased over time by \shorttilde110.8\% and \shorttilde57.8\% respectively.

\subsection{Execution Graphs}
\label{section:execution_graphs}
We next evaluate the merging, batching and branching capabilities of the SGX-Bundler library.

\subsubsection{Merging}
\label{section:result-merging}


\img[0.5]{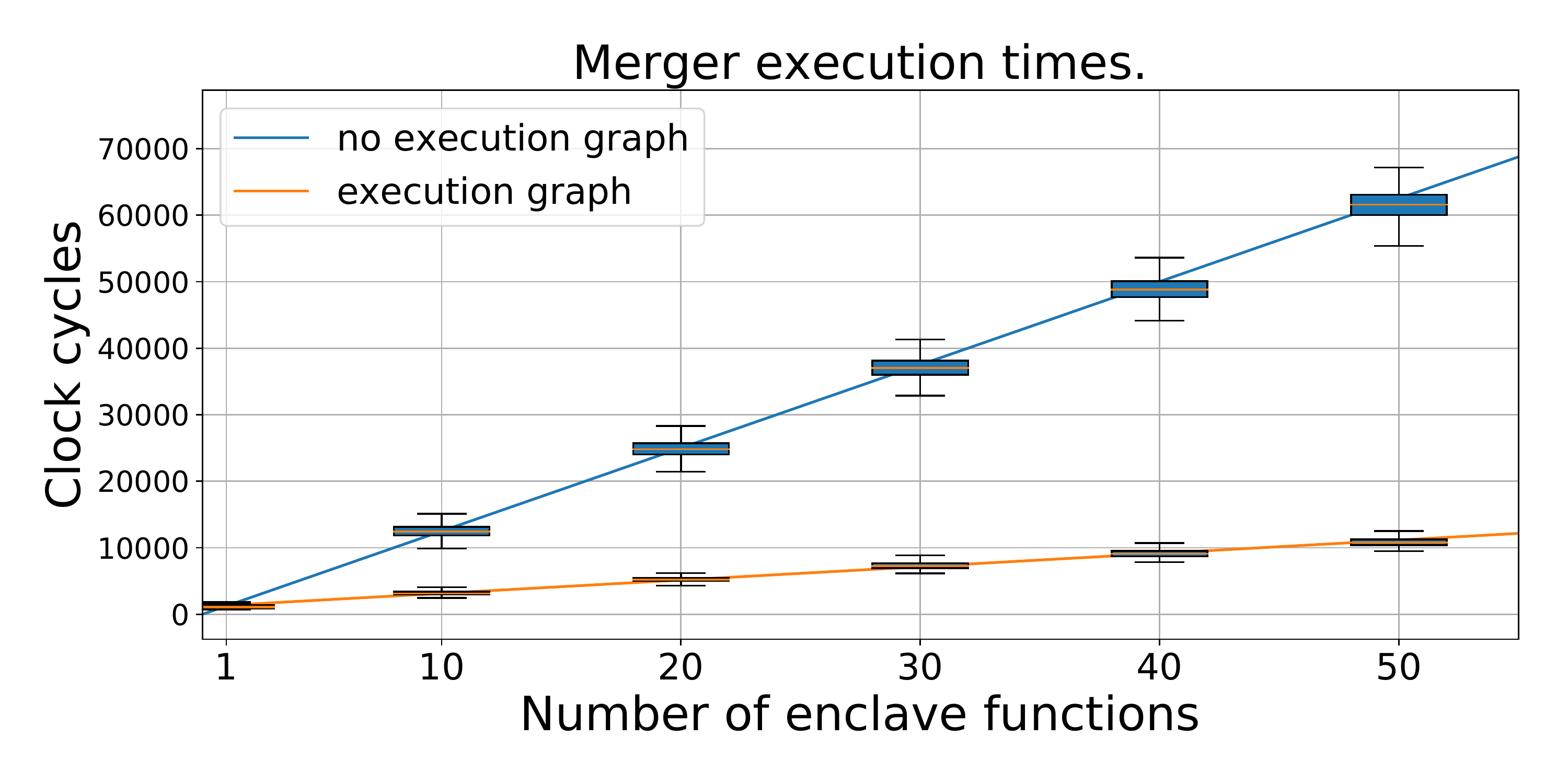}{figure:merging}{Execution times when executing $n$ different enclave functions, with and without execution graphs.}


We compared execution time of merging \newline $n \in \{1, 10, 20, 30, 40, 50\}$ enclave function calls using execution graphs to single enclave function calls (i.e. no merging). As Figure~\ref{figure:merging} illustrates, merging significantly reduces execution time when multiple calls are performed. Note that without execution graphs the execution time is dominated by the transition time measured in Figure~\ref{figure:hotcall_transition}.

\subsubsection{Batching}
\label{section:result:batch}

Given a list of $n \in \{1, 10, 20, 30, 40, 50\}$ elements, we compared the processing time using either a loop or an iterator. As Figure~\ref{figure:graph_iter_loop} illustrates, iterators are \shorttilde14x faster even though both operations grow linearly with the number of elements.

Currently, iterators are limited to a single enclave call per round while loops can perform an arbitrary number. 
We measured the execution time of applying $m\in \{1, 5, 10, 15, 20\}$ enclave functions to an input list of size 20. 
Despite iterating through the list $m$ times, iterators are still \shorttilde 6.4 times faster than loops (see Figure~\ref{figure:graph_iter_loop_2}).

\imgstacked{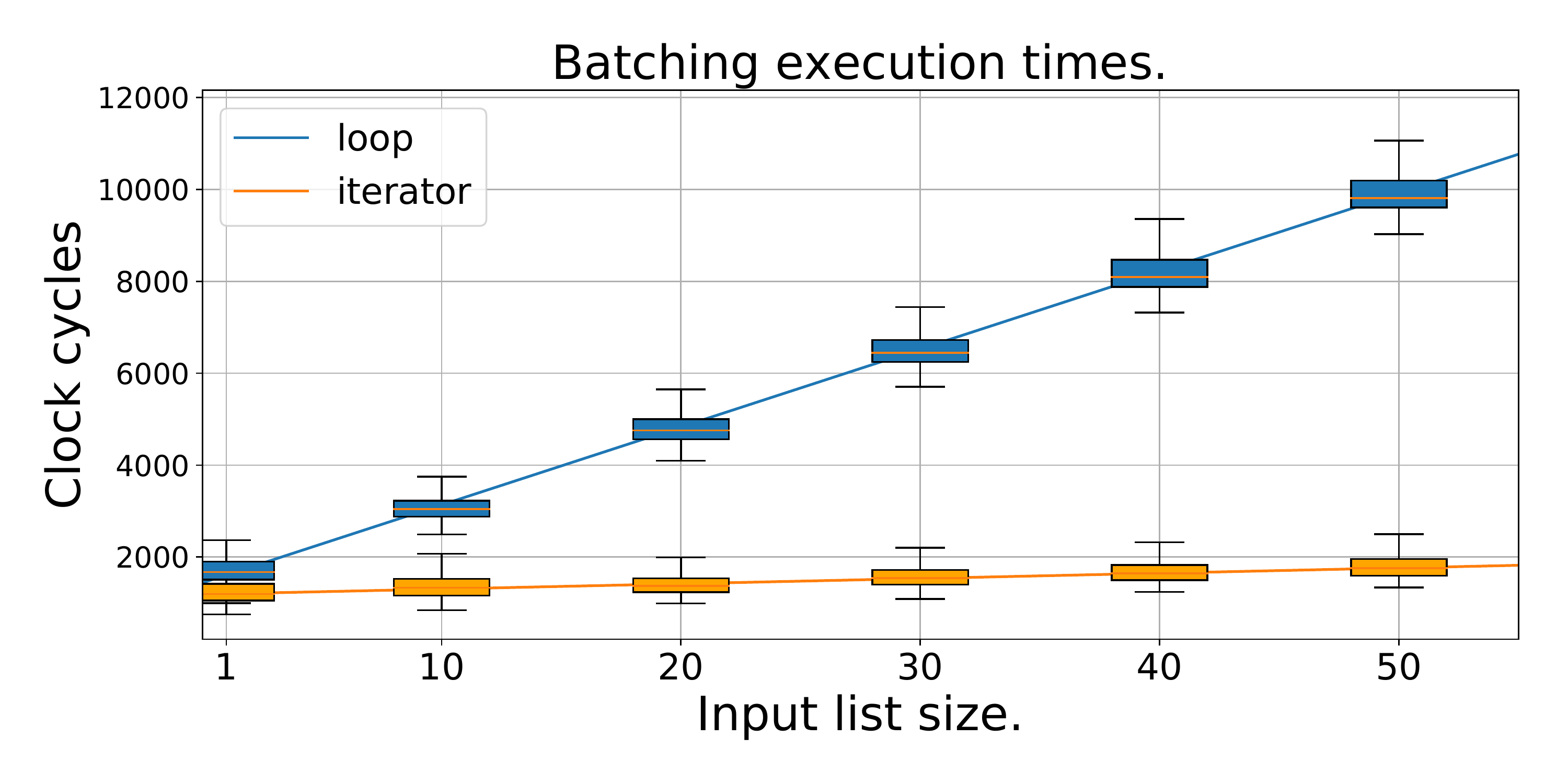}{One function, $n$ elements.}{figure:graph_iter_loop}
{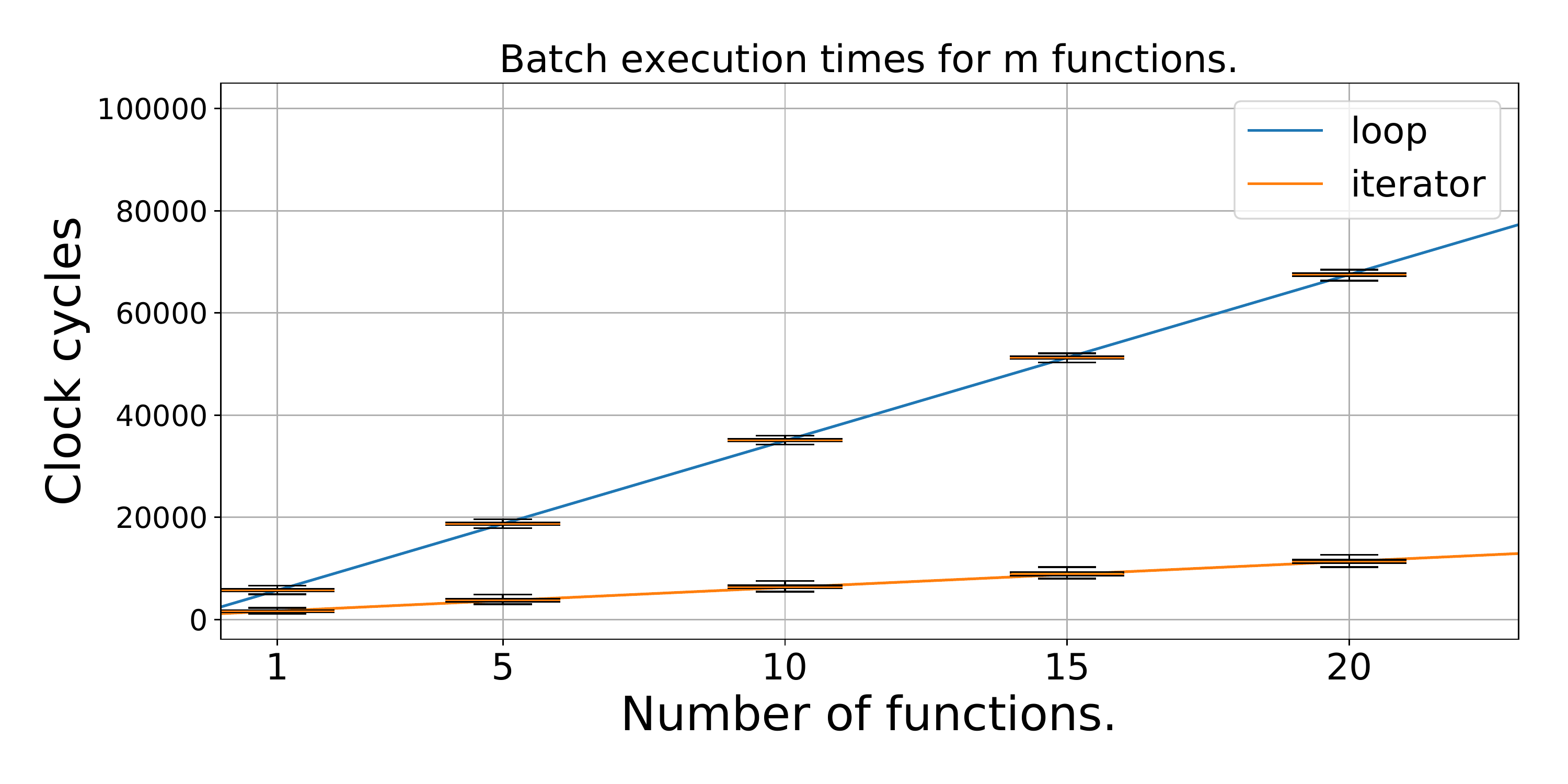}{$m$ functions, $20$ elements.}{figure:graph_iter_loop_2}
{Execution time when applying $m$ enclave functions to each element in a list of size $n$.}{figure:for_merge}

\subsubsection{Branching}
\label{section:result:branching}

\begin{figure}[htb]
	\centering
	\begin{subfigure}[b]{0.45\textwidth}
		\includegraphics[width=1\linewidth]{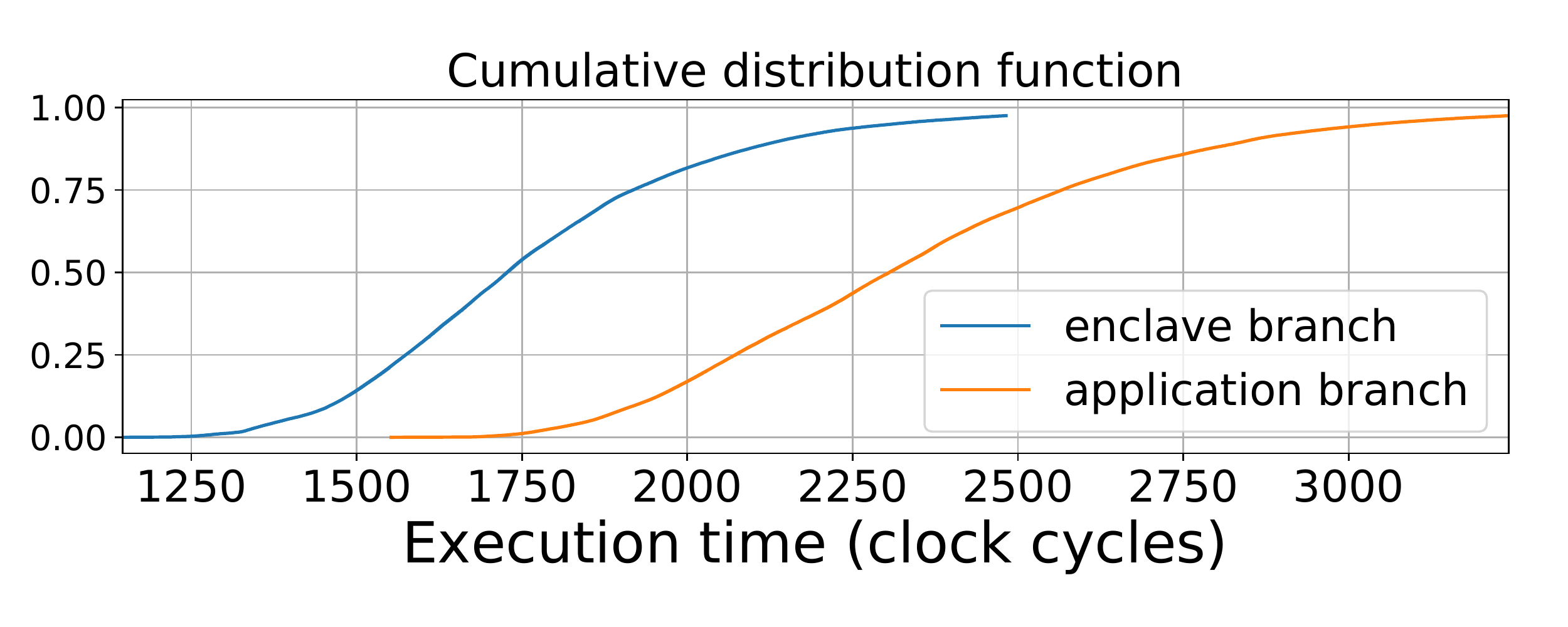}
		\caption{True condition}
		\label{figure:enclave_branching:sub1}
	\end{subfigure}
	\begin{subfigure}[b]{0.45\textwidth}
		\centering
		\includegraphics[width=1\linewidth]{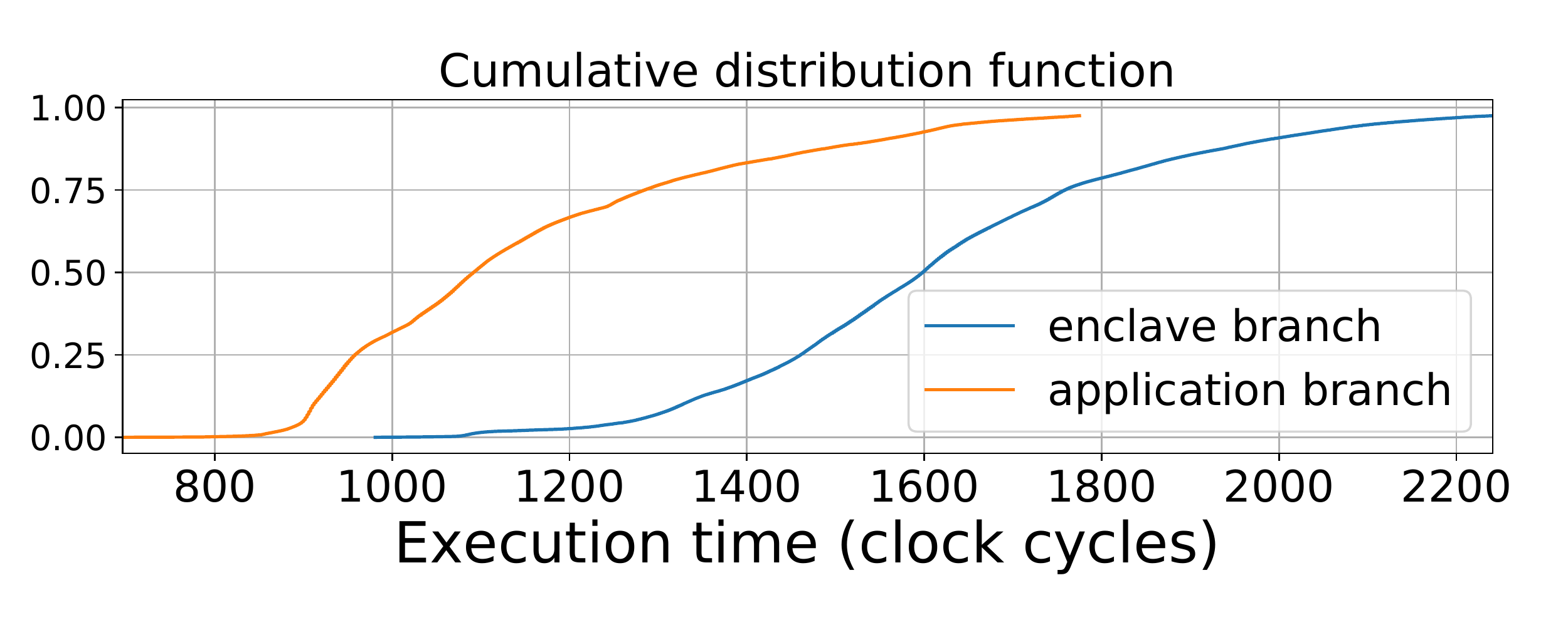}
		\caption{False condition}
		\label{figure:enclave_branching:sub2}
	\end{subfigure}
	\begin{subfigure}[b]{0.45\textwidth}	
		\centering
		\includegraphics[width=1\linewidth]{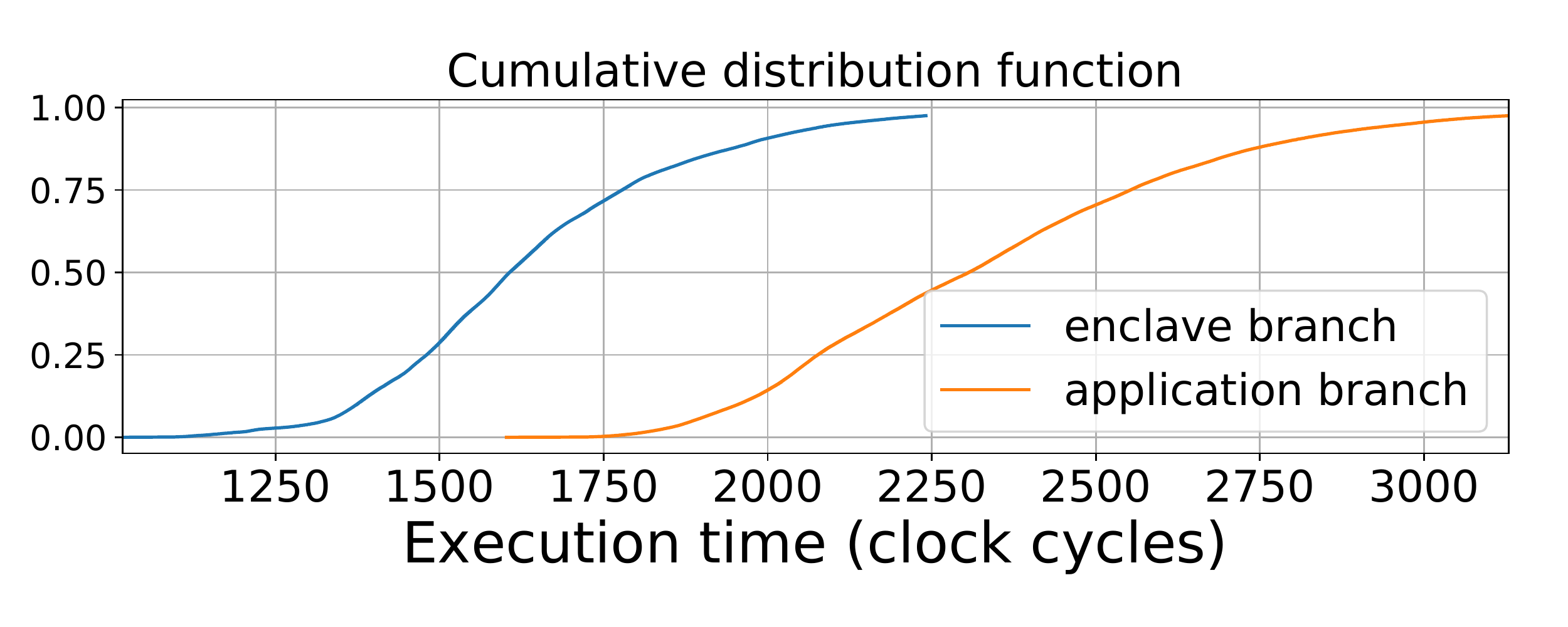}
		\caption{False condition in loop with 2 iterations}
		\label{figure:enclave_branching:sub3}
	\end{subfigure}
	\vspace{-0.3em}
	\caption{Execution times for enclave and application branching.}
	\label{figure:enclave_branching}
\end{figure}

In Figure~\ref{figure:enclave_branching} we compare enclave branching, which happens inside the enclave to application branching which performs the branch in the application and executes the branch body as a separate execution graph. Notice that enclave branching is faster when the branch condition is true, but slower when false. However, if the branch operation is repeated at least once (Figure~\ref{figure:enclave_branching} (c)), then enclave branching is faster even when the condition is false.

\subsubsection{Enclave Function Memoization}
\label{section:result:memo}

We measure performance gains of memoizing a variable accessed through an enclave function. In Figure~\ref{figure:memoize_1} we measure execution time for a cache hit with LRU or FIFO eviction policies. Notice that a cache hit is \shorttilde20-24x faster than a cache miss.

At the same time, memoization introduces a noticeable overhead to enclave operations that update a value and thus must update or invalidate the cached value. Figure~\ref{figure:memoize_2} shows that the cost differs depending on the state of the cache (\shorttilde12.0\% and \shorttilde20.7\% for warm and cold caches respectively).
\imgstacked{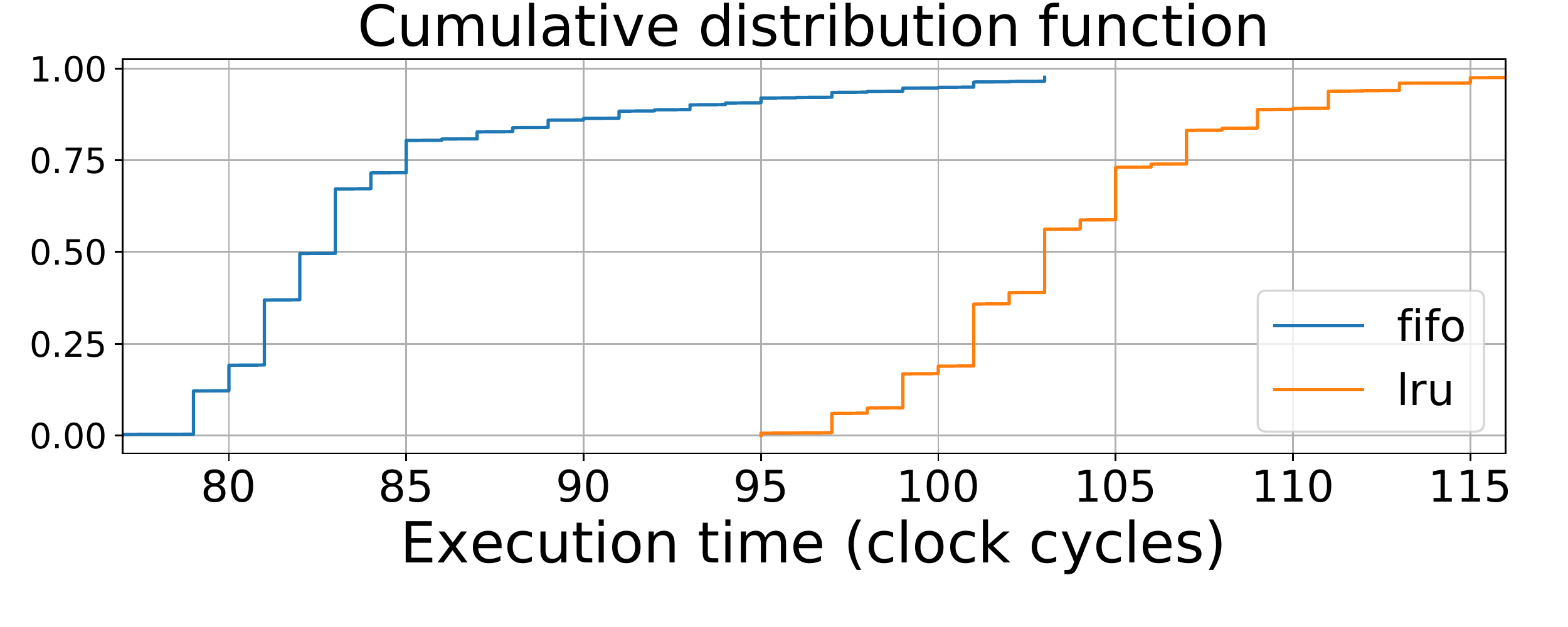}{Cache hit}{fig:memo:sub1}{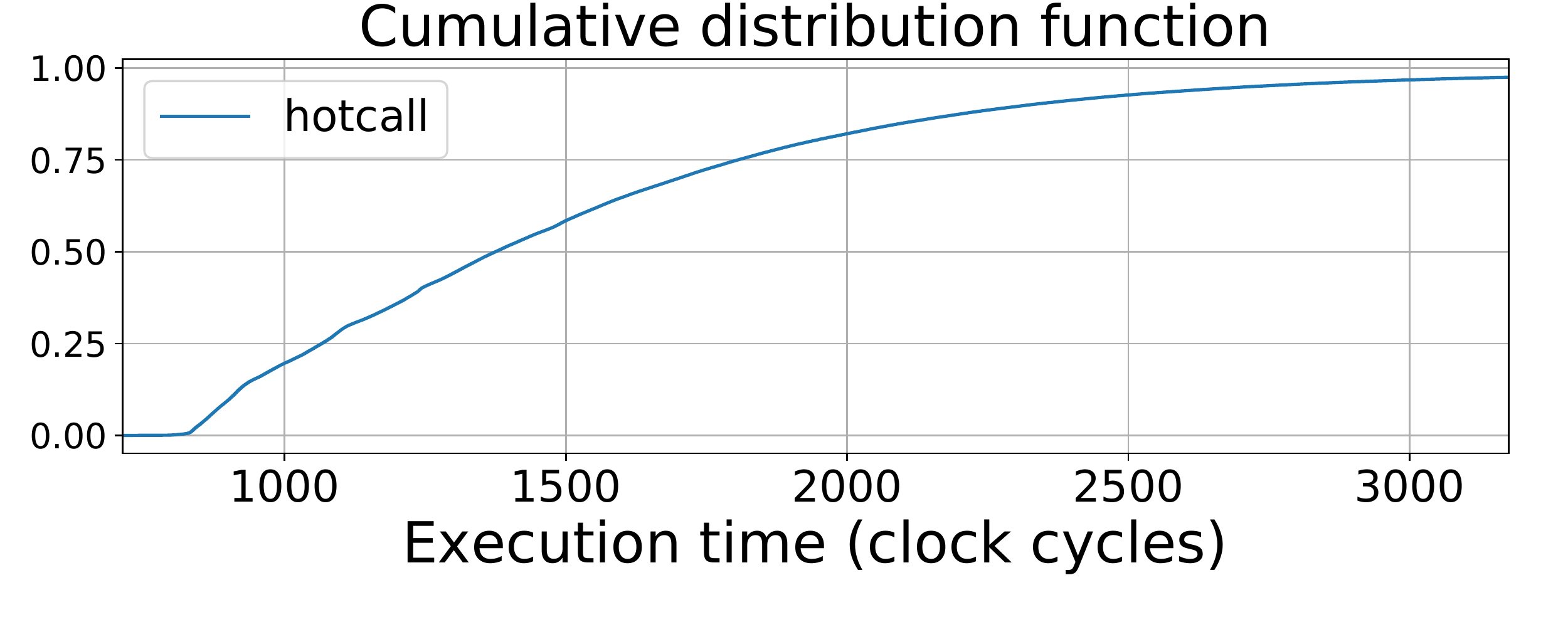}{Cache miss}{fig:memo:sub2}{Execution times of an enclave function with memoization enabled for both cache hits and misses.}{figure:memoize_1}


\img[0.5]{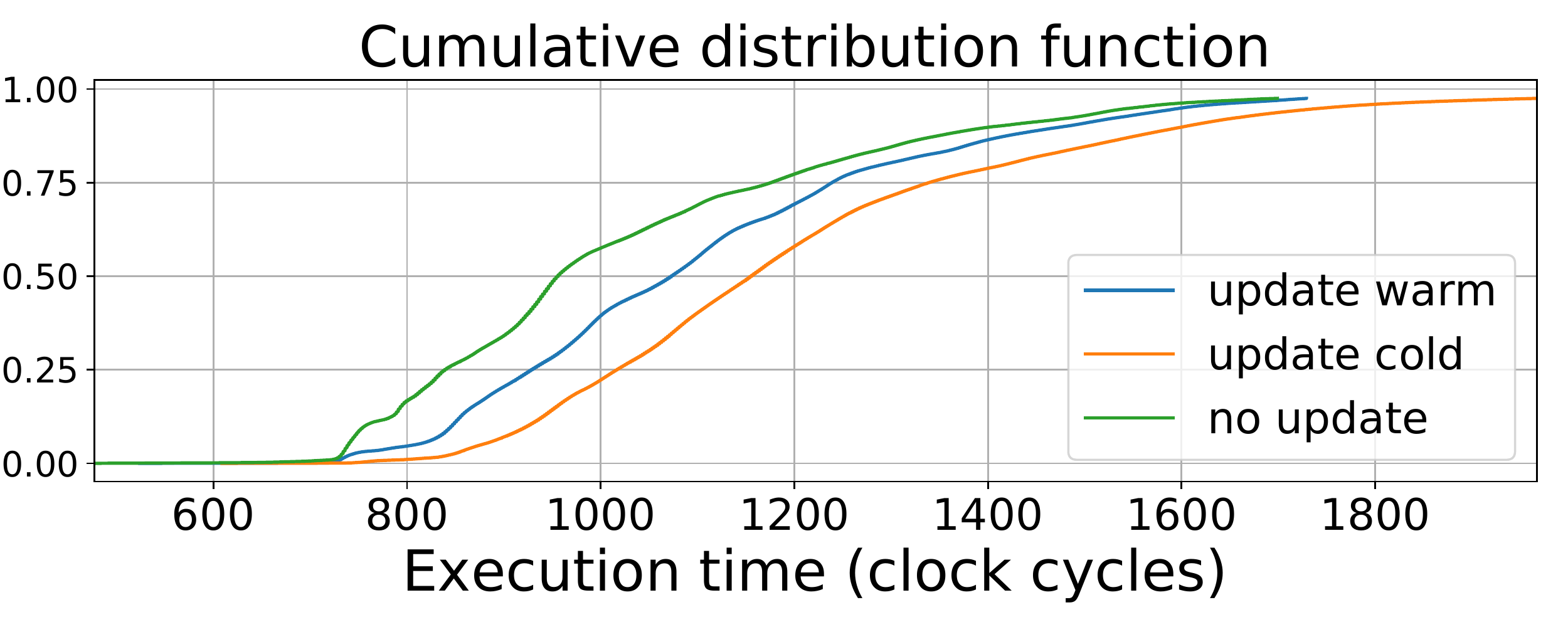}{figure:memoize_2}{Enclave function execution times with and without cache updates.} 


\subsection{OvS Prototypes}
\label{section:results-ovs}

To better assess the proposed solutions it is important to study the performance impact in non-trivial real-world applications. Since time constraints would limit us to analysis of a single application, we chose Open vSwitch (commit \texttt{53cc4b0}) where different operations and workloads should cover a wide range of executions characteristics.
To this aim, we studied four Open vSwitch flow table operations (add, delete, modify and evict flow rule) under realistic SDN workloads and as average of 20 separate rounds.

The performance of each operation has been compared across five different implementations: \textit{baseline} is the original version,
\textit{SGX vanilla} is the OFTinSGX version~\cite{medina:2019}, 
\textit{Switchless} uses hotcalls instead of ECALLs~ \cite{Weisse2017}
while \textit{Bundler} uses all optimizations described in this paper.
Finally \textit{SGX refactored} is the authors heavily modified version tailored specifically for SGX and will be used to compare the trade-offs between performance and development effort. 
All evaluation scripts are openly available\footnote{Source code repository:~\url{https://github.com/nicopal/sgx_bundler}}.


\subsubsection{Add and Delete Flow Rules}

\imgstacked{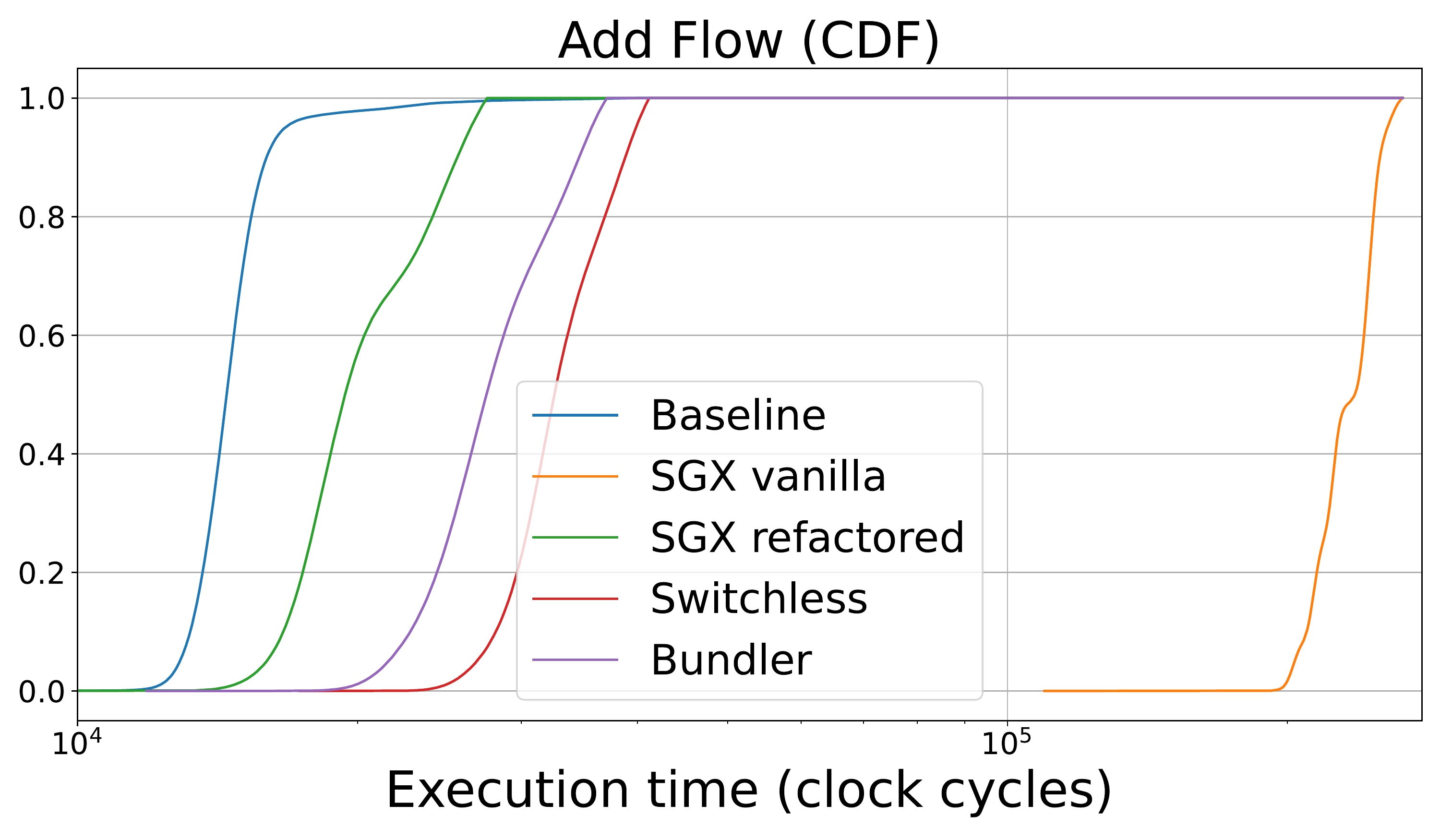}{Add flow
}{figure:ovs_add}{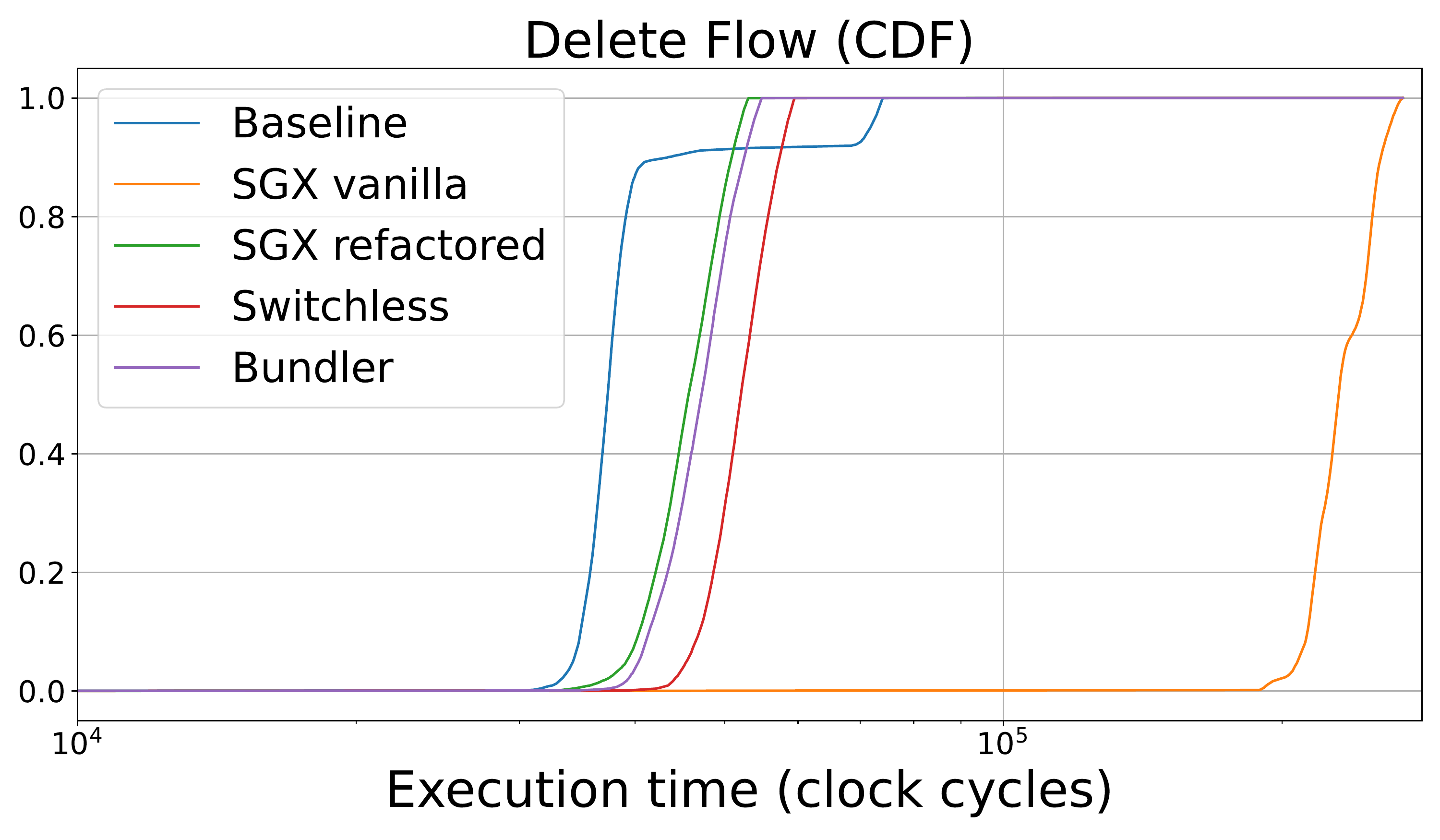}{Delete flow}{figure:ovs_del}{Execution time for add and delete flow operations}{figure:add-delete}

Measured execution time for add and delete flow operations can be seen in Figure \ref{figure:add-delete} and Table~\ref{table:overhead}.
We note that in both cases \textit{SGX vanilla} performs significantly worse than other implementations. It is also noted that \textit{Bundler} performs slightly better than \textit{Switchless} but not as well as \textit{SGX refactored} which is the best preforming secure implementation (with \textit{Baseline} being the best performing overall implementation.)

The difference between \textit{Bundler} and \textit{Switchless} increases slightly in the case of delete operations. Unlike add, delete operations often target multiple table entries and therefore benefits from batching. We will shortly revisit this difference for other operations

\imgstacked{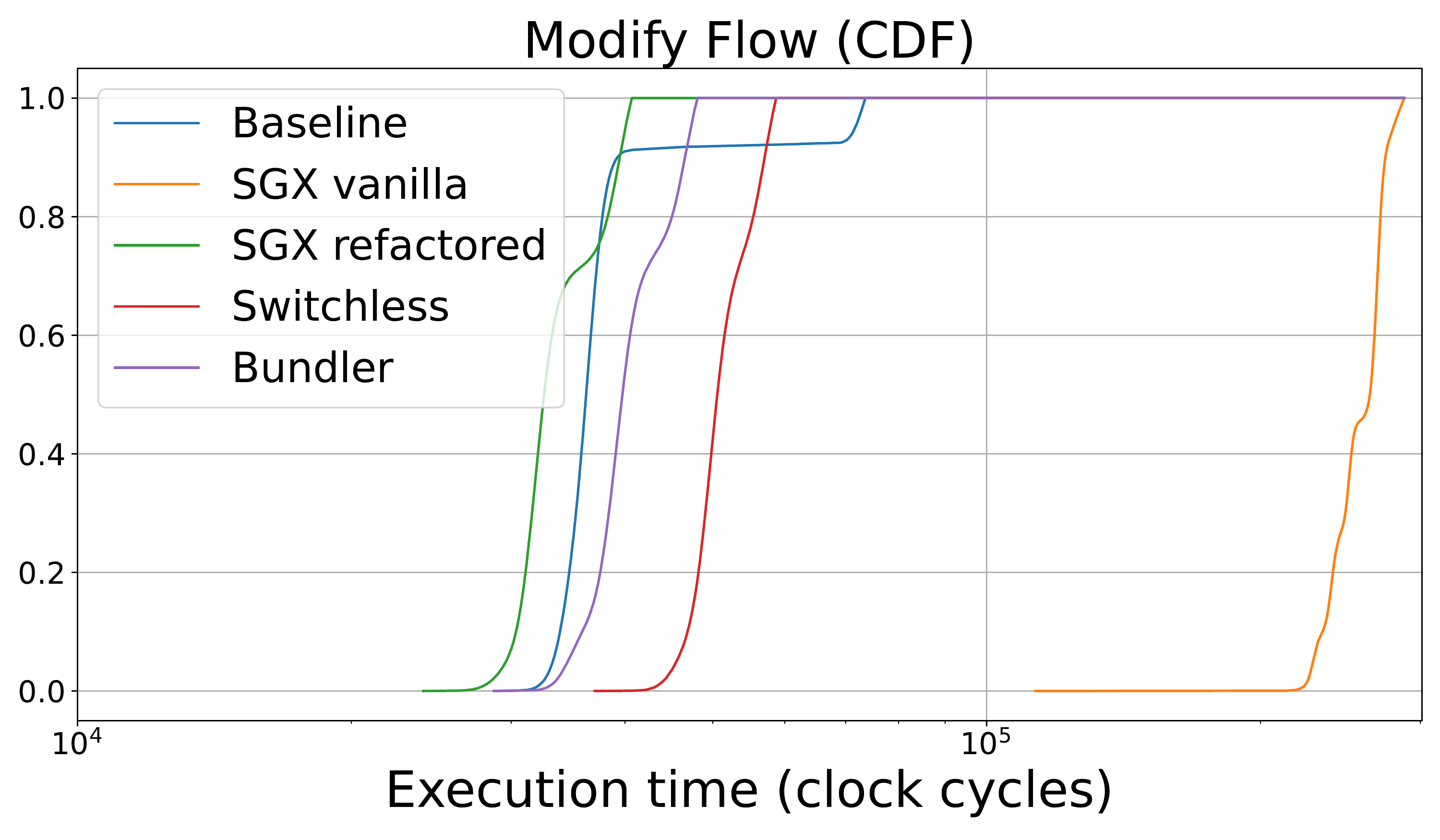}{Modify flow
}{figure:ovs_mod}{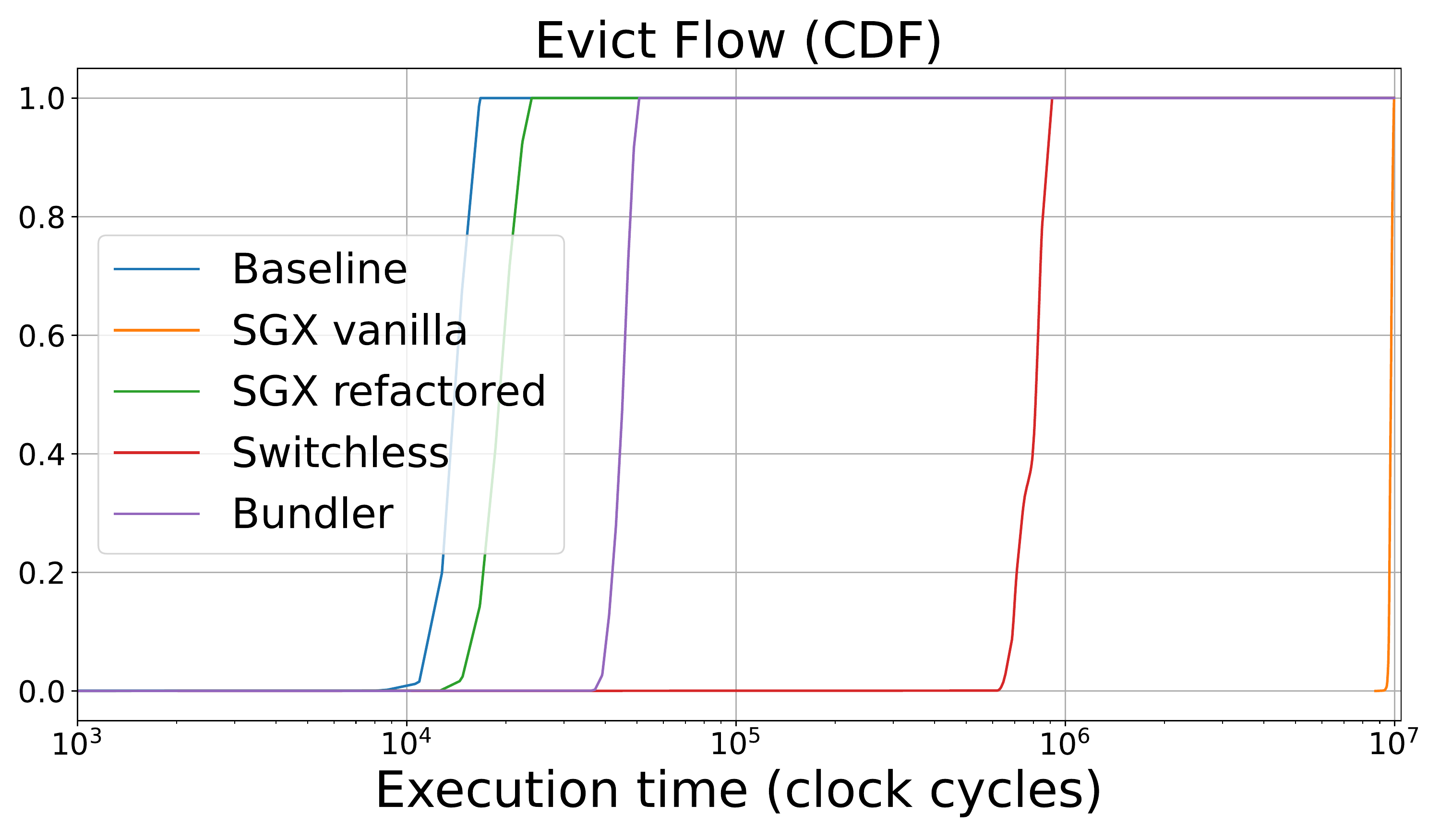}{Evict flow}{figure:ovs_evict}{Execution time for modify and evict flow operations}{figure:mod-evict}

\subsubsection{Modify and Evict Flow Rules}

Measured execution time for modify and evict operations can be seen in Figure \ref{figure:mod-evict} and  Table~\ref{table:overhead}. The pattern observed earlier is repeated here, although the performance difference between \textit{Bundler} and \textit{Switchless} is growing. For example, for evict operations the former is well over an order of magnitude faster.

We note that modify and evict represent more complex operations that may require many more enclave transitions. In such situations \textit{Bundler} appears to provide much more stable performance improvements. Hence we conclude that while \textit{Switchless} provides some performance improvements, its benefits are limited in IO-intensive applications.

\begin{table*}[htb]
  \centering
  \caption{OvS overhead for different operations}
  \label{table:overhead}
  \begin{small}
  \begin{tabular}{|l|c||c|c|c|c||c|c|c|c|c}
    \hline
    
    \textbf{Version} &
	\textbf{Batch} &
    \textbf{Op} &    
    \multicolumn{3}{c||}{\textbf{Overhead for quantile}} & 
	\textbf{Op} &    
    \multicolumn{3}{c|}{\textbf{Overhead for quantile}} \\
	
    & \textbf{size} & 
	& \textbf{25\%} & \textbf{50\%} & \textbf{75\%} &
	& \textbf{25\%} & \textbf{50\%} & \textbf{75\%} \\

    \hline
    \hline
    Baseline & - & Add & 0 & 0 & 0 & Delete & 0 & 0 & 0 \\
    SGX Refactor & - & Add & 29 & 34 & 53 & Delete & 19 & 22 & 26 \\
    Bundler & 1 & Add & 80 & 90 & 106 & Delete & 22 & 26 & 29 \\
	Bundler & 16 & - & - & - & - & Delete & 17 & 16 & 15 \\
    Switchless & - & Add & 119 & 124 & 136 & Delete & 37 & 39 & 42 \\
    SGX vanilla & - & Add & 1473 & 1524 & 1511 & Delete & 508 & 517 & 544 \\
    \hline
    Baseline & - & Modify & 0 & 0 & 0 & Evict & 0 & 0 & 0 \\
    SGX Refactor & - & Modify & -11 & -10 & 0 & Evict & 23 & 25 & 27 \\
    Bundler & 1 & Modify & 8 & 10 & 16 & Evict & 153 & 153 & 150 \\
    Bundler & 16 & Modify & 21 & 21 & 21 & Evict & 86 & 89 & 91 \\
    Switchless & - & Modify & 38 & 39 & 45 & Evict & 3661 & 3906 & 3860 \\
    SGX vanilla & - & Modify & 588 & 622 & 615 & Evict & 49628 & 47606 & 45703 \\    
    \hline
  \end{tabular}
\end{small}
\end{table*}

Given these results, it seems that while \textit{Bundler} introduce a measurable performance overhead it does not drastically increase execution time even in corner cases. We note that \textit{SGX refactored} demonstrates better performance, but is the performance gain worth the amount of work required to rewrite the original application?

\subsection{Programming Effort Trade-Offs}
The effort required to rewrite and optimize an application specifically for SGX depends on the size and complexity of the application. In practice doing this may not be possible due to cost and time limitations, lack of know-how or other issues. For example, maintainability may suffer as transferring new changes from the original project to the rewritten version becomes much harder. As mainline changes also include security patches, this might also negatively affect the security.

While in this work the authors were able to create the \textit{SGX refactored} implementation for Open vSwitch, the effort for doing so was not negligible. In comparison the \textit{Bundler} implementation utilized the SGX-Bundler library and required much smaller changes to the Open vSwitch source code (less than 1\% of lines and 1\% of files were modified).
We believe this is mainly attributed to the user API for constructing execution graphs (see Appendix~\ref{appendix:api}).

\subsection{Security Analysis and Limitations}
\label{section:bundle-limitation}
We next assess the security implications of using SGX-Bundler by examining the changes that can affect the trusted code running inside the enclave.

\noindent
\textbf{Storing execution graphs in shared memory} does not advantage an attacker in the Intel SGX adversary model, where the attacker controls the underlying OS and the enclave IO~\cite{anati:2013}.
SGX-Bundler applies the approach employed by Hotcalls~\cite{Weisse2017} and SGX SDK (reference) for passing the data structures between the untrusted code and the enclave.
Furthermore, execution graphs are initially constructed in the untrusted, rich execution environment and are not sensitive to attacks on confidentiality and integrity (beyond performance effects). 
We further refer to the security analysis in~\cite{Weisse2017} which analyzes a similar approach. 

\noindent
\textbf{Replay attacks and Denial of Service:}
The SGX-Bundler approach does not introduce additional risks for replay attacks or denial of service.
Calls to the API of the trusted application running in the enclave are always issued from the untrusted rich execution environment.
Denial of service is outside the Intel SGX adversary model.

\noindent
\textbf{Confidentiality of application data:}
SGX-Bundler does not affect data confidentiality as neither user nor application data is stored in the shared memory.
We assume that all user data communicated between the trusted enclave and the untrusted rich execution environment is done through a secure channel established following enclave attestation~\cite{anati:2013}.

\noindent
\textbf{Integrity of the memoization cache:} 
An attacker can temporarily change the content of a memoization cache without being detected.
If an attacker modifies a memoization cache entry and restores the original value before the next memoization cache verification, then the unauthorized modification will not be detected by the enclave.
An attacker can time the cache changes such that cache modifications are removed before each cache verification, thus hiding the changes.
However, assuming the attacker cannot determine exactly when the enclave worker thread verifies memoization caches, the integrity violation will be eventually detected. 
As a result, enclave function memoization only guarantees \textit{eventual integrity} of its content and is not directly suitable for settings requiring stronger integrity guarantees.
Instead, it is appropriate in contexts when the logic of the application must be protected, and not the data that it processes. Consider the case of  proprietary algorithms, software implementations, parameters of DNN models, and other types of software intellectual property.




\section{Related Work} 
\label{section:related-work}
Performance issues in SGX applications can sometimes be attributed to the high cost of entering and exiting enclaves.
Weisse \etal introduced "\textit{HotCalls}" for communicating with enclaves using shared untrusted memory \cite{Weisse2017}.
This approach can be orders of magnitude faster than ECALLs, although the use of untrusted memory also increases the attack surface for the enclave. The switchless enclave function call component of the \textit{SGX-Bundler} library developed in this paper, presented in Section \ref{section:bundler}, is heavily inspired by this work.

The HotCalls protocol requires an enclave worker thread that communicates with the main thread through a shared memory region. This thread will occupy one CPU core, which is economical only when the SGX enclave is under some load. Tian \etal suggested using an adaptive approach where ECALLs are used when the device is mostly idle and switchless calls are used when it is under some load \cite{Tian2018}. This scheme has been included in recent versions of the Intel SGX SDK as an official feature. 
We chose to not use this scheme in our paper as it lacked the flexibility and control granularity of a custom solution. 

\textit{ShieldStore} is an SGX enabled key-value store that uses HotCalls and also overcomes EPC memory limitations by storing all key-value pairs in encrypted untrusted memory \cite{Kim2019}. The two prototypes developed in this work were highly influenced by ShieldStore.

The authors of~\cite{DinhNgoc2019} presented an extensive performance study for virtualization and Intel SGX. 
The study includes a large number of benchmarks on ECALL, OCALL, and EPC paging performance in native and virtual environments. The native ECALL performance estimates were used in Section \ref{section:results-ovs} of this paper.

Weichbrodt \etal presented \textit{sgx-perf}, a performance analysis tool for Intel SGX applications \cite{Weichbrodt2018}.
Using this tool, the authors analyzed scenarios where Intel SGX was a significant performance bottleneck and suggested possible solutions. 
Two such scenarios were subsequent calls of the same enclave function and subsequent calls to different enclave functions.
While the authors proposed batching and merging respectively or moving the caller into the enclave, to the best of our knowledge this has not been implemented and described before our work.
Furthermore, we improve the usability of this approach by packaging it as a library.

Software-Defined Networking (SDN) and in particular the SDN control plane has been extensively scrutinized by security researchers \cite{Shu2016},\cite{Abdou2018}.
Some researchers have considered the use of Trusted Computing and trusted execution to address security issues. 
For example, Jacquin \etal proposed using TPM to ensure a trusted boot and use of attestation to monitor the integrity of flow tables \cite{Jacquin2015}, while Paladi \etal suggested using Intel SGX to ensure a secure boot and to provide secure communication channels \cite{paladi2017}. Similarly, Shih \etal proposed executing parts of a virtual network function inside an Intel SGX enclave \cite{10.1145/2876019.2876032}.

Medina \etal proposed \textit{OFTinSGX}, an Open vSwitch implementation where OpenFlow flow tables are placed inside an SGX enclave \cite{medina:2019}. While this provided confidentiality and integrity guarantees to the flow tables, it also brought a significant performance degradation to OvS. 

\section{Conclusions}
\label{section:conclusions}
 
In this paper we presented SGX-Bundler, a mechanism to help improve the performance of IO-intensive applications in Intel SGX enclaves.
The proposed mechanism combines switchless SGX communication and a novel optimization using execution graphs and function memoization. 
We extended earlier work and developed two prototypes that utilize switchless communication both with and without execution graphs and memoization. 
Our evaluation shows that while switchless communication contributes to some performance improvements, the addition of execution graphs and memoization leads to further significant improvements. In particular, our approach seems to be much better equipped to handle  exceedingly IO-intensive operations, making it more suitable for real-world usage.
The suggested improvements come however at the cost of increased size and complexity. To facilitate adoption we encapsulated the proposed mechanism into the openly available SGX-Bundler library, thereby reducing development effort significantly.

We thoroughly evaluated the performance improvements introduced by the SGX-Bundler library using the case study of Open vSwitch, a widely used network switch implementation.
The SGX-Bundler library can be used for other IO-intensive applications that can benefit from the security guarantees of isolated execution in SGX, such as biological sequence analysis or long-running simulations.
Considering the many parameters that each evaluation entails, we will explore in future work the performance effects of the SGX-Bundler library in further applications, as well as evaluate the required programming efforts across several case studies.

%
%
%
\bibliographystyle{splncs04}
\bibliography{paper}

\newpage

\begin{appendices}
\renewcommand{\thesection}{\Alph{section}}%

\section{Execution Graphs}
\label{appendix:executiongraphs}

\begin{figure}[th]
	\centering
	\includegraphics[width=0.49\textwidth]{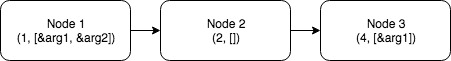}
	\caption{A simple execution graph consisting of three enclave function calls.}
	\label{figure:simple_graph}
\end{figure}

Execution graphs are represented as an array of items where each entry is either an enclave function, a control statement, or an iterator. In its simplest form, an execution graph is only a list of enclave functions that are executed sequentially. This enables arbitrary merging of enclave function calls and fulfills functional requirement (2) listed in Section~\ref{section:implementation-func_req}. 
Figure~\ref{figure:simple_graph} illustrates a simple execution graph with three enclave function calls. Each function call is represented by a tuple (\textit{function\_id}, \textit{function\_data}), where \textit{function\_data} is a list of pointers. By convention, the last element in the function\_data list is the address where potential return values are written.

We next describe other node types that can be used to construct execution graphs. 
These enable batching and branching and fulfill the functional requirements 3 and 4 of Section \ref{section:implementation-func_req}.

\subsubsection{Iterator}
\label{section:implementation-iterator}

Iterator-nodes allow multiple invocations of an enclave function with different parameters in a single enclave transition. Iterators can be used to implement functional style operators such as \textit{map} and \textit{for-each}. 

\begin{figure}[th]
	\centering
	\includegraphics[width=0.49\textwidth]{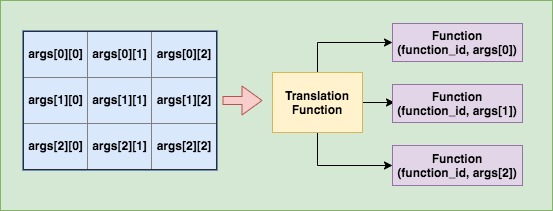}
	\caption{Iterator implementation.}
	\label{figure:iter_node}
\end{figure}

Figure~\ref{figure:iter_node} illustrates the high-level overview of iterator implementation. 
Note that input parameters are now represented as a matrix which will be processed by a translation function before each row is used for one invocation of the target function.

\subsubsection{If}
\label{section:implementation-if}

\begin{figure}[th]
	\centering
	\includegraphics[width=0.4\textwidth]{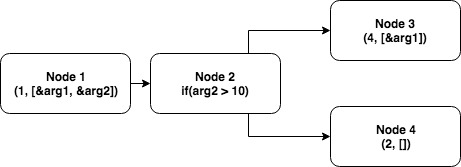}
	\caption{Execution graph containing an if-node.}
	\label{figure:if_node}
\end{figure}

If-nodes choose between two possible execution paths, possibly depending on the result of a previous enclave operations.
An example is shown in Figure~\ref{figure:if_node} where the choice between two enclave functions depends on value of \texttt{arg2}, which may have been modified in the previous enclave function call.
The implementation allows use of complex conditions in postfix notation~\cite{Hamblin1962} and automatically handles required type conversions.

\subsubsection{For}
\label{section:implementation-for}

\begin{figure}[th]
	\centering
	\includegraphics[width=0.5\textwidth]{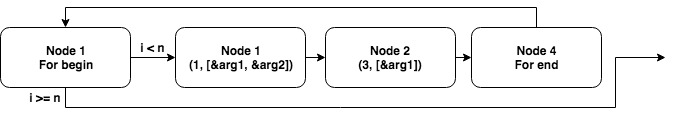}
	\caption{Execution graph containing a for-node.}
	\label{figure:for_node}
\end{figure}

A \textit{for}-node allows to execute a subset of the execution graph repeatedly for $n$ iterations.
This is an alternative to iterators presented in Section~\ref{section:implementation-iterator}.
However, for-loops can execute an arbitrary subset of the execution graph in each iteration while an iterator can only execute a single enclave function. 
Figure~\ref{figure:for_node} illustrates a execution graph containing a for-node. 
Note that a for-loop requires two additional nodes to be inserted in the execution graph. One node in the front and one in the end. 
All parameters of enclave functions in the for-loop body which are marked as list parameters are automatically incremented in each loop iteration.

\subsubsection{While}
\label{section:implementation:while}
A while-node allows to execute a subset of the execution graph repeatedly conditioned on a boolean expression. 
A while-node is implemented in the same way as the for-node presented earlier. 
The difference between for and while-nodes is the loop condition. 
While-nodes use a boolean condition as loop condition which is implemented the same way as the boolean conditions in if-nodes.

\section{SGX-Bundler User API}
\label{appendix:bundler-api}
\label{appendix:api}

The SGX-Bundler library consists in practice of two libraries for the untrusted and trusted halves of the application. Furthermore, a set of preprocessor macros are provided to simplify development and will be discussed later in this appendix.

\begin{table*}[t!]
    \small
	\begin{tabular}{ | p{5.5cm} | p{12cm} |}
		\hline
		\textbf{Function} & \textbf{Description} \\
		\hline
		\texttt{void hotcall\_init(struct shared\_memory\_ctx *sm\_ctx, 
		sgx\_enclave\_id\_t eid);} 
		& Starts a new enclave worker thread in the enclave with id \textit{eid} using the shared memory region defined by \textit{sm\_ctx}; spawns a new thread that calls \textit{ecall\_thread\_init} listed in Table \ref{table:trusted_api}. \\
		\hline
		\texttt{void hotcall\_destroy(struct shared\_memory\_ctx *sm\_ctx);} 
		& Destroys the enclave worker thread which is associated with the shared memory region pointed to by \textit{sm\_ctx}.\\
		\hline
	\end{tabular}
	\caption{Public functions of the untrusted library component.}
	\label{table:untrusted_api}
\end{table*}

\begin{table*}[t!]
    \small
	\begin{tabular}{ | p{6cm} | p{11cm} |}
		\hline
		\textbf{Function} & \textbf{Description} \\
		\hline
		\texttt{int ecall\_thread\_init(struct shared\_memory\_ctx *sm\_ctx);}
		& Contains a spinlock loop which will loop indefinitely,  polling the shared memory region pointed to by \textit{sm\_ctx}. Must be called from a child thread since it will not return until destroyed explicitly by calling \textit{hotcall\_destroy} listed in Table \ref{table:untrusted_api}. This function is an ECALL and is called from the untrusted application. \\
		\hline
		\texttt{void
		hotcall\_register\_config( struct hotcall\_config *config);}
		& 
		Configures the trusted part of the library. 
		Struct \textit{hotcall\_config} contains the mapping between function ids and enclave functions. This function is called from the enclave.
		\\
		\hline
	\end{tabular}
	\caption{Public functions of the trusted library component.}
	\label{table:trusted_api}
\end{table*}

The untrusted and trusted parts of the SGX-Bundler library only expose two functions each, listed in tables \ref{table:untrusted_api} and \ref{table:trusted_api}, respectively. The untrusted part further also exposes a large number of function calls for constructing execution graphs. 
These methods are not meant to be used directly by users and are further described in our technical report\footnote{Reference removed during submission}.

The user API simplifies construction of execution graphs and can be used with 
in both imperative and functional fashion.

\subsection{Integration with Trusted Applications}
\label{section:implementation-integration}

Integration of the SGX-Bundler library requires minimal modifications to trusted and untrusted applications. 
The required modifications are described below.

\begin{enumerate}
	\item Link the trusted and untrusted part of the SGX-Bundler library to the trusted and untrusted part of the SGX application, respectively. 
	\item Assign a function id to each enclave function that needs to be accessible from the untrusted application. 
	The \texttt{function\_id} is used by both the trusted and untrusted part of the application. 
	\item Define translation functions in the enclave for each enclave function of the enclave's API, as described in Section \ref{section:wrapper}.

	\item Create a mapping between function ids and corresponding translation function. This is done by defining an array of pointers large enough to fit all enclave function calls.
	\item Register the \textit{function\_id} to translation function mapping with the trusted part of the library using function \textit{hotcall\_register}, presented in Table \ref{table:trusted_api}.
	\item Launch an enclave worker thread from the untrusted part of the application using the \textit{hotcall\_init} function, presented in Table \ref{table:untrusted_api}.
\end{enumerate}

Following the steps enumerated above is enough to integrate the SGX-Bundler library in an Intel SGX application and start an enclave worker thread ready to process switchless enclave function calls.
The final step of the integration process consists of replacing all ECALL invocations in the source code files of the untrusted application with switchless enclave function calls, using the user API presented in the following.

\subsection{Calling a Switchless Enclave Function}

Listing \ref{listing:hotcall} demonstrates invocation of the enclave function \textit{hotcall\_plus\_one}. This operation is blocking until the enclave thread has executed and any returns values have been retrieved.

\begin{lstlisting}[
    basicstyle=\small, 
    language=C, 
    label={listing:hotcall},
    caption=]
    int x = 0;
    HCALL(
        CONFIG(.function_id = hotcall_plus_one), 
        VAR(x, 'd'));
    if(x != 1) printf(``Wrong answer\n");
\end{lstlisting}

\subsection{Merging Enclave Functions with Execution Graphs}

Listing \ref{listing:merge} demonstrates merging multiple enclave functions.
After BUNDLE\_BEGIN() the HCALL statements create new nodes in a execution graph that is submitted for execution only when BUNDLE\_END() is reached.

\begin{lstlisting}[
    basicstyle=\small, 
    language=C, 
    label={listing:merge},
    caption=
]
    int x = 0;
    BUNDLE_BEGIN();
    HCALL(
        CONFIG(.function_id = hotcall_plus_one), 
        VAR(x, 'd'));
    HCALL(
        CONFIG(.function_id = hotcall_plus_one), 
        VAR(x, 'd'));
    BUNDLE_END();
    if(x != 2) printf("Wrong answer\n");
\end{lstlisting}

\subsection{For Loop}

Listing \ref{listing:for} demonstrates use of a for-loop to construct an execution graph to increment elements of the list \textit{xs} by 1.

\begin{lstlisting}[
    basicstyle=\small, 
    language=C, 
    label={listing:for},
    caption=
]
    unsigned int n_iters = 10;
    int xs[n_iters] = { 0 };
    BUNDLE_BEGIN();
    BEGIN_FOR(((struct for_config) {
        .n_iters = &n_iters}));
        HCALL(
            CONFIG(.function_id = hotcall_plus_one), 
            VECTOR(xs, 'd'));
    END_FOR();
    BUNDLE_END();
    
    for(int i = 0; i < n_iters; ++i) {
        if(xs[i] != 1) printf("Wrong answer\n");
    }
\end{lstlisting}

\subsection{For Each}

Listing \ref{listing:for_each} demonstrates creation of an execution graph using for-each to  increment members of a list by 1 using an iterator.

\begin{lstlisting}[
    basicstyle=\small, 
    language=C, 
    label={listing:for_each},
    caption=
]
    unsigned int n_iters = 10;
    int xs[n_iters] = { 0 };
    BUNDLE_BEGIN();
    FOR_EACH(((struct for_each_config) { 
            .function_id = hotcall_plus_one, 
            .n_iters = &n_iters}), 
        VECTOR(xs, 'd')
    );
    BUNDLE_END();
    
    for(int i = 0; i < n_iters; ++i) {
        if(xs[i] != 1) printf("Wrong answer\n");
    }
\end{lstlisting}

\subsection{Map}

Listing \ref{listing:map} demonstrates use of the map operator to increment elements in a list. Note that the output is written to ys and the input list xs is not modified.

\begin{lstlisting}[
    basicstyle=\small, 
    language=C, 
    label={listing:map},
    caption=]
    unsigned int n_iters = 10;
    int xs[n_iters] = { 0 }, ys[n_iters] = { 0 };
    BUNDLE_BEGIN();
    MAP(
        ((struct map_config) {
            .function_id = hotcall_plus_one_ret, 
            .n_iters = &n_iters }),
        VECTOR(xs, 'd'),
        VECTOR(ys, 'd')
    );
    BUNDLE_END();
    
    for(int i = 0; i < n\_iters; ++i) {
        if(xs[i] != 0) printf("Wrong answer\n");
        if(ys[i] != 1) printf("Wrong answer\n");
    }
\end{lstlisting}

\subsection{If}

Listing \ref{listing:if} demonstrates using a an if operator to implement execution graph branching.


\begin{lstlisting}[
    basicstyle=\footnotesize, 
    language=C, 
    label={listing:if},
    caption=
]
    BUNDLE_BEGIN();
        int x = 5, y = 7;
        IF(((struct if_config) { .predicate_fmt = "d>d" }),
            VAR(y, 'd'), VAR(x, 'd')
        ); THEN HCALL(
                CONFIG(.function_id = hotcall_plus_one), 
                VAR(x, 'd'));
        END
    BUNDLE_END();
    if(x != 8) printf("Wrong answer\n"); 
\end{lstlisting}

\subsection{Other API components}
The user API provides additional functionality not shown here due to space limitations. This includes other operators such as while-loops and support for enclave function memoization.

\end{appendices}

\end{document}